\documentclass{cernrep} 
\usepackage{texnames}
\usepackage[T1]{fontenc}
\begin{document}
\title{Particle-in-Cell Codes for plasma-based particle acceleration}
 
\author{Alexander Pukhov}

\institute{University of Dusseldorf, 40225 Dusseldorf, Germany}

\maketitle 

\begin{abstract}
Basic principles of particle-in-cell (PIC ) codes with the main application for plasma-based acceleration are discussed. The ab initio full electromagnetic relativistic PIC codes provide the most reliable description of plasmas. Their properties are considered in detail. Representing  the most fundamental model, the full PIC codes are computationally expensive. The plasma-based acceleration is a multi-scale problem with very disparate scales. The smallest scale is the laser or plasma wavelength (from one to hundred microns) and the largest scale is the acceleration distance (from a few centimeters to meters or even kilometers). The Lorentz-boost technique allows to reduce the scale disparity at the costs of complicating the simulations and causing unphysical numerical instabilities in the code. Another possibility is to use the quasi-static approximation where the disparate scales are separated analytically.
\end{abstract} 

\section{Introduction}

Plasma-based particle acceleration involves a rather nonlinear medium,
the relativistic plasmas \cite{Pukhov}. This medium requires proper
numerical simulation tools. During the past decades, Particle-in-Cell
(PIC) methods have been proven to be a very reliable and successful
method of kinetic plasma simulations \cite{Buneman,Dawson,Langdon,Hockney}.
This success of PIC codes relies to a large extent on the very suggestive
analogy with the actual plasma. The plasma in reality is an ensemble
of many individual particles, electrons and ions, interacting with
each other by the self-consistently generated fields. The PIC code
is very similar to that, with the difference that number of the numerical
particles, or macroparticles we follow in the code, may be significantly
smaller. One may think as if one numerical ``macroparticle'' is
a clump, or cloud, of many real particles, which occupy a finite volume
in space and all move together with the same velocity. The consequent
conclusion is that we have a ``numerical plasma'' consisting of
heavy macroparticles, which have the same charge-to-mass ratio as
the real plasma electrons and ions, but substitute many of those.

Depending on the application, different approximations can be chosen.
The most fundamental approximation is the full electromagnetic PIC
codes solving the Maxwell equations together with the relativistic
equations of motion for the numerical particles. These ``ab initio``
simulations produce the most detailed results, but can be also very
expensive. In the case of long scale acceleration, when the driver
propagates distances many times larger than its own length, the quasi-static
approximation can be exploited. In this case, it is assumed that
the driver changes little as it propagates distance comparable with
its own length. The quasi-statics allows for separation of fast and
slow variables and great acceleration of the simulation at the cost
of radiation: the laser pulse or any emitted radiation cannot be described
directly by such codes. Rather, a an additional module for the laser
pulse is required, usually in the envelope approximation.

In this work, we describe the basic principles of the PIC methods,
both full electromagnetic and quasi-static in application to plasma-based
acceleration.

\section{The basic equations}

First, let us formulate the problem we are going to solve. We are
doing electromagnetic and kinetic simulations. This means, we are
solving the full set of Maxwell equations \cite{Jackson}

\begin{eqnarray}
\frac{\partial{\bf E}}{\partial t} & = & c\nabla\times{\bf B}-4\pi{\bf j},\label{Ampere}\\
\frac{\partial {\bf B}}{\partial t} & = & -c\nabla\times{\bf E},\label{Faraday}\\
\nabla\cdot{\bf E} & = & 4\pi\rho,\label{Poisson}\\
\nabla\cdot{\bf B} & = & 0,\label{Bcharge}
\end{eqnarray}

\noindent where we use CGS units and $c$ is the speed of light in
vacuum. 

Let us stop for a moment at this very fundamental system of equations.
The electric and magnetic fields, ${\bf E,B}$, evolve according to
the time-dependent equations (\ref{Ampere})-(\ref{Faraday}) with
the source term in the form of current density ${\bf j}$. This current
is produced by the self-consistent charge motion in our system of
particles. It is well known from the textbooks on electrodynamics
(see, e.g., \cite{Jackson}), that the Gauss law~(\ref{Poisson})
together with the curl-free part of Eq.~(\ref{Faraday}), lead to
the charge continuity equation

\begin{equation}
\frac{\partial \rho}{\partial t}+\nabla\cdot{\bf j}=0.\label{cont}
\end{equation}

\noindent One can apply the  operator $\nabla\cdot$ to
the Faraday law (\ref{Faraday}) and use the Gauss Eq.~(\ref{Poisson})
for $\nabla\cdot{\bf E}$, to obtain (\ref{cont}). The opposite is
true as well. If the charge density always satisfies the continuity
Eq.~(\ref{cont}), the Gauss Eq.~(\ref{cont}) is fulfilled automatically
during evolution of the system, if it was satisfied initially. The
symmetric consideration is valid for the magnetic field ${\bf B}$.
As there is no magnetic charges, Eq.~(\ref{Bcharge}) remains always
valid, if it was valid initially.

This means, we may reduce our problem to a solution of the two evolutionary
equations (\ref{Ampere})-(\ref{Faraday}) considering Eqs.~(\ref{Poisson}),(\ref{Bcharge})
as initial conditions only. This appears to be a very important and
fruitful approach. PIC codes using it have a ``local'' algorithm,
i.e., at each time step the information is exchanged between neighboring
grid cells only. No global information exchange is possible because
the Maxwell equations have ``absolute future'' and ``absolute past''
\cite{Landau}. This property make the corresponding PIC codes perfectly
suitable for parallelizing and, in addition, the influence (of always
unphysical) boundary conditions is strongly reduced.

\section{Kinetics and hydrodynamics}

Now we have to define our source term, ${\bf j}$. In general, for
this purpose, we have to know the distribution function of plasma
particles

\begin{equation}
F^{N}({\bf x_{1},p_{1},...,x_{N},p_{N}}).\label{FN}
\end{equation}

\noindent It defines the probability of an $N-$particles system to
take a particular configuration in the $6N-$dimensional phase space.
Here ${\bf x_{n},p_{n}}$ are coordinates and momenta of the $n$-th
particle. The function (\ref{FN}) provides \textit{the exhaustive}
description of the system. However, as it is shown in statistical
physics (see, e.g., \cite{Klimontovich}), single-particle distribution
function for each species of particles may be sufficient to describe
the full system. The sufficient condition is that the inter-particle
correlations are small, and can be treated perturbatively. The equation,
which governs the evolution of the single-particle distribution function
$f({\bf x,p})$ is called Boltzmann-Vlasov Eq. \cite{Braginskii,Vlasov}:

\begin{equation}
\frac{\partial f}{\partial {t}}+\frac{{\bf p}}{m\gamma}\nabla f+\frac{{\bf F}}{m}\nabla_{p}f=St,\label{BV}
\end{equation}

\noindent where $m$ is the single-particle mass of the corresponding
species, $\gamma=\sqrt{1+(p/mc)^{2}}$ is the relativistic factor,
${\bf F}$ is the force, and $St$ is the collisional term (inter-particle
correlations).

The kinetic equation (\ref{BV}) on the single-particle distribution
function is $6-$dimensional and still complicated. It is a challenge
to solve it analytically or numerically.

It appears, however, that under appropriate conditions one can make
further simplifications. It is shown in statistics, that inter-particle
collisions lead to the Maxwellian distribution function ($H-$theorem,
see, e.g., \cite{Klimontovich}). In the non-relativistic case the
Maxwellian distribution has the form:

\begin{equation}
f({\bf {x,v})=\frac{n({\bf x)}}{\sqrt{2\pi T}}e^{-\frac{({\bf v-V})^{2}}{2mT}},}
\end{equation}

\noindent where $n({\bf x})$ is the local particle density, $T$
is the temperature, and ${\bf V}$ is the local streaming velocity.
The characteristic time for establishing of the Maxwellian distribution
is the inter-particle collision time.

Thus, if the effective collisional time in the considered system is
short in comparison with other characteristic times, the distribution
function remains Maxwellian. It is sufficient then to write evolutionary
equations on the \textit{momenta} of the distribution function, such
as the local density

\begin{equation}
n({\bf x})=\int\, f({\bf x,v})\, d^{3}v{\bf \label{n}}
\end{equation}

\noindent the hydrodynamic velocity

\begin{equation}
{\bf V}({\bf x})=\int\,{\bf v}f({\bf x,v})\,{\bf dv}{\bf ,\label{V}}
\end{equation}

\noindent and the temperature

\begin{equation}
T({\bf x})=\int\,\frac{({\bf v-V})^{2}}{2m}f({\bf x,v})\, d^{3}v{\bf ,\label{T}}
\end{equation}

\noindent 
are called fluid-like, or of hydrodynamic type.

\section{Vlasov and Particle-in-Cell codes}

If, however, the distribution function does deviate, or is expected
to deviate significantly from the Maxwellian one, we have to solve
the Boltzmann-Vlasov Eq.~(\ref{BV}). What could be the appropriate
numerical approach here?

\begin{figure}
\centerline{\includegraphics[clip,width=0.9\textwidth]{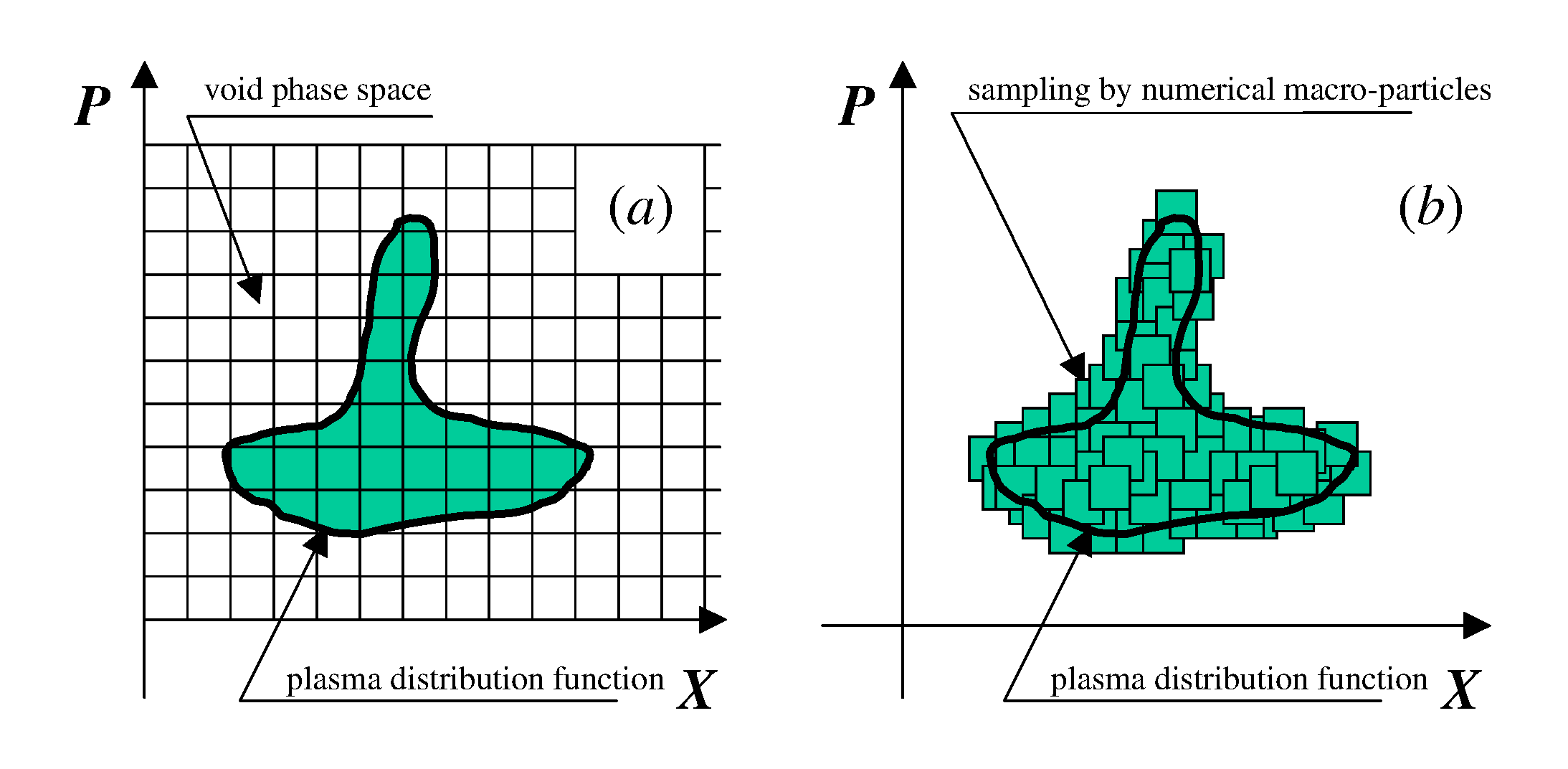}} \protect\caption{Kinetic plasma simulations. (a) Vlasov method: Eulerian grid in the
phase space; (b) PIC method: numerical macroparticles mark the distribution
function.}

\label{Vlasov:fig} 
\end{figure}


At the first look, it seems that the most straightforward and relatively
simple approach is to solve the partial differential Eq.~(\ref{BV})
using finite differences on the Eulerian grid in phase space. Indeed,
this approach is pursued by several groups \cite{Ruhl}, and gains
even more popularity as the power of available computers grows. One
of the potential advantages of these \textit{Vlasov} codes is the
possibility of producing ``smooth'' results. Indeed, the Vlasov
codes handle the distribution function, which is a smoothly changing
real number already giving the probability to find the plasma particles
in the corresponding point of the phase space.

The Vlasov codes, however, are very expensive from the computational
point of view, and even one-dimensional problems may demand the use
of parallel super-computers. The reason, why these codes need so much
computational power, becomes clear from Fig.~\ref{Vlasov:fig}a.
It shows schematically a mesh one would need for a ${\bf 1d1v}$ Vlasov
code. The notation ${\bf 1d1v}$ means, that the code resolves one
spatial coordinate and one coordinate in the momentum (velocity) space.
The dashed region is to represent the part of the phase space occupied
by plasma particles, where the associated $2-$dimensional distribution
function $f(x,p_{x})$ is essentially non-zero. The unshaded region
is empty from particles, and nothing interesting happens there. Nevertheless,
one has to maintain these empty regions as parts of the numerical
arrays, and process them when solving Eq.~(\ref{BV}) on the Eulerian
grid. This processing of empty regions leads to enormous wasting of
computational power. This decisive drawback becomes even bolder with
increase in the dimensionality of the problem under consideration.
The efficiency of Vlasov codes drops exponentially with the number
of dimensions, and becomes miniscule in the real ${\bf 3d3v}$ case,
when one has to maintain in the memory and process a $6-$dimensional
mesh, most of it just empty from particles.

Now let us show that there is another, presently more computationally
effective, method to solve the Boltzmann-Vlasov Eq.~(\ref{BV}).
This is the \textit{finite-element} method. The principle is illustrated
in Fig.~\ref{Vlasov:fig}b. Again, imagine some distribution function
in the phase space (the shaded region). Now, let us approximate, or
sample, this distribution function by a set of Finite Phase-Fluid
Elements (FPFE):

\begin{equation}
f({\bf x,p})=\sum_{n}~W_{n}^{ph}~S^{ph}({\bf x-x}_{n},{\bf p-p}_{n})\label{sample}
\end{equation}

\noindent where $W_{n}^{ph}$ is the ``weight'' of the FPFE, and
$S^{ph}({\bf x,p})$ is the ``phase shape'', or the support function
in the phase space. The center of the FPFE is positioned at ${\bf x}_{n},~{\bf p}_{n}$.
We are free to make a particular choice of the support function. For
simplicity, we choose here a ${\bf 6d}$ hypercube:

\[
S^{ph}({\bf x},{\bf p})=1,~~|x_{\alpha}|<\frac{\Delta x_{\alpha}}{2},~~|p_{\alpha}|<\frac{\Delta p_{\alpha}}{2},~~\alpha=x,y,z
\]

\noindent where $\Delta x_{\alpha}$ is the FPFE size along the $j-$axis
in configuration space, and $\Delta p_{\alpha}$ is the FPFE size
along the $p_{\alpha}-$axis in momenta space.

The ``phase fluid'' transports the distribution function along the
characteristics of the Boltzmann-Vlasov Eq.~(\ref{BV}) (see, e.g.,
\cite{Chen}). Thus, we have to advance the centers of FPFE along
the characteristics:

\begin{eqnarray}
\frac{d {\bf x}_{n}}{d t} & = & \frac{{\bf p}}{m\gamma},\label{char_x}\\
\frac{d {\bf p}_{n}}{d t} & = & {\bf F+F_{st}},\label{char_p}
\end{eqnarray}

\noindent where ${\bf F_{st}}$ is the effective ``collisional''
force due to the collisional term in Eq.~(\ref{BV}). The FPFE follow
the evolution of distribution function in the phase space. Of course,
the Eqs.~(\ref{char_x})-(\ref{char_p}) are the relativistic equations
of motion of particles! Thus, the FPFE method is equivalent to the
PIC method.

The significant advantage of the finite element method over the Vlasov
codes is that one does not need to maintain a grid in the full phase
space. Instead, our FPFE sample (or mark) only the interesting regions,
where particles are present, and something important is going on.
We still do maintain a grid in the \textit{configuration} space to
solve the field equations (\ref{Ampere})-(\ref{Faraday}), but this
grid has only 3 (and not 6 as in the Vlasov case) dimensions. Thus,
PIC codes may be considered as ``packed'', or ``Lagrangian'' Vlasov
codes. Moreover, the finite phase fluid element approach is even more
fundamental than the Boltzmann-Vlasov equation itself, because it
may be easily generalized to the case when one macroparticle corresponds
to just one real particle, and when inter-particle correlations are
not small. The corresponding codes are usually called P$^{3}$M: particle-particle-particle-mesh
codes \cite{Hockney}.

As soon as we consider our macroparticles not as simply ``large clumps
of real particles'', but as finite elements in the phase space, we
discover, that there is no fundamental obstacle to the simulation
of a cold plasma. Moreover, it is in this case where the finite elements
approach becomes effective computationally and superior to the Vlasov
codes. The phase space of a cold plasma is degenerate. The particles
occupy a mere $3-$dimensional hypersurface of the full $6-$dimensional
phase space. Evidently, this hypersurface can be accurately sampled
even by a relatively small number of macroparticles (FPFE). As the
system evolves, this surface deforms, stretches and contracts, but
remains degenerate and three-dimensional, unless any heating (diffusion
in the phase space) is present. There is a full stock of interesting
physical phenomena in relativistic laser-plasma interactions one would
like to study, where the physical collisional heating is negligible.
Unfortunately, the numerical heating present in the ``standard''
PIC codes \cite{Langdon} leads to an unphysical numerical diffusion
in the phase space, which spoils the picture. The code able to simulate
initially cold plasma must be energy conservative.

\section{Continuity equation}

Historically, the first particle-in-cell codes were electrostatic
\cite{Hockney}, and they have to solve explicitly the Poisson equation

\begin{equation}
\nabla^{2}\phi=-4\pi\rho,\label{Poisson_phi}
\end{equation}

\noindent giving the static electric field

\begin{equation}
{\bf E}_{\parallel}=-\nabla\phi.\label{Estat}
\end{equation}

\noindent The way of generalization to the electromagnetic case seemed
quite natural. One simply adds the vector-potential ${\bf A}$:

\begin{equation}
{\bf E}={\bf E}_{\parallel}+{\bf E}_{\perp}=-\nabla\phi-\frac{1}{c}\frac{\partial{\bf A}}{\partial t}.\label{Efull}
\end{equation}

Yet, there is another way of treating the electromagnetic fields.
We have established in the previous section, that the simultaneous
solution of Ampere's law (\ref{Ampere}) and the continuity equation
(\ref{cont}) satisfies the Gauss law~(\ref{Poisson}) automatically.
Thus, one may work with the fields ${\bf E}$ and ${\bf B}$ directly,
not introducing the electrostatic potential $\phi$, and not solving
the Poisson equation (\ref{Poisson_phi}).

\begin{figure}
\centerline{\includegraphics[clip,width=0.5\textwidth]{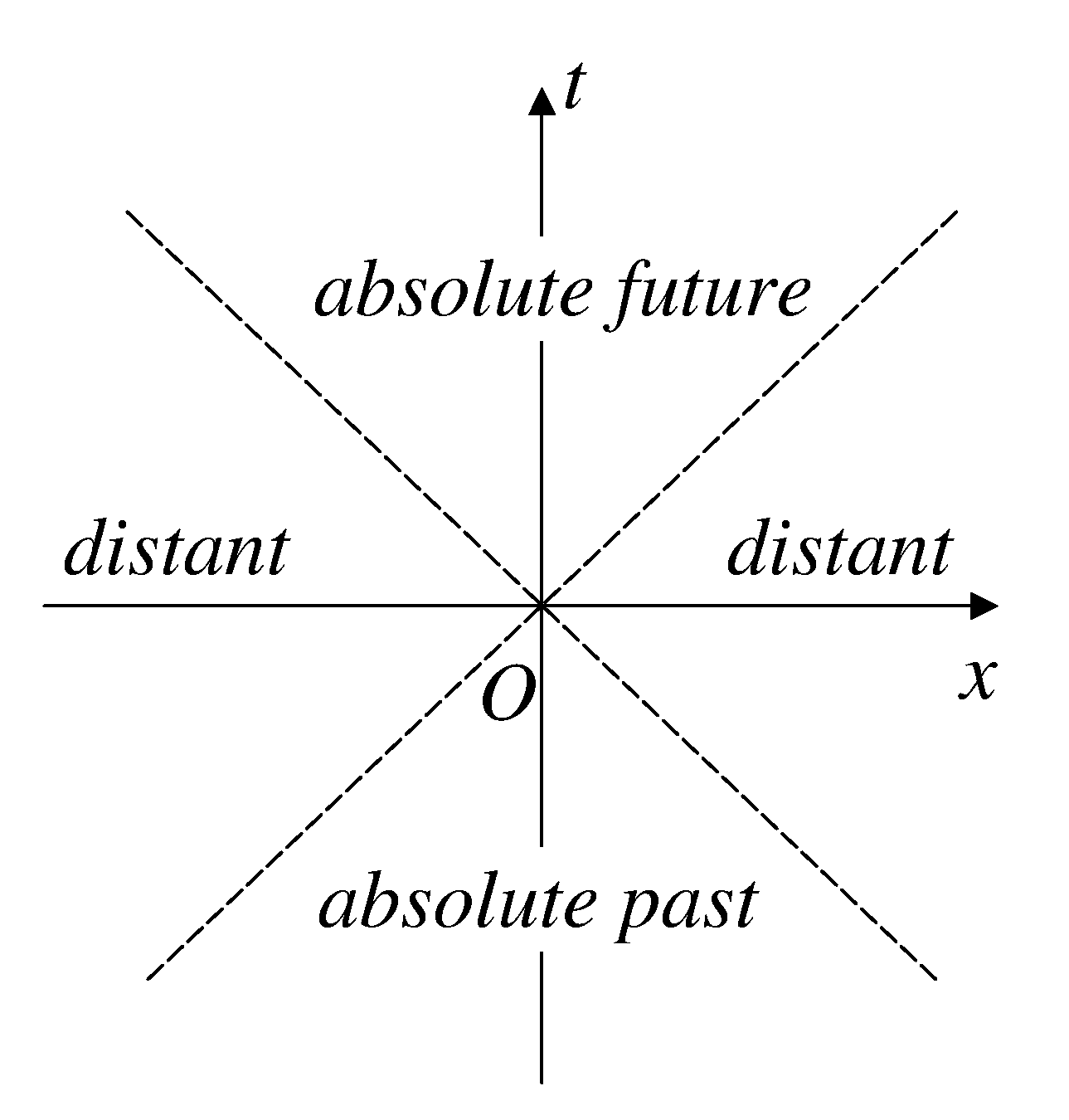}} \protect\caption{Minkovski space-time diagram illustrating the causality in the special relativity theory.}

\label{fig:Minkovski} 
\end{figure}


It is very advantageous to avoid solution of the Poisson equation
(\ref{Poisson_phi}), as it is \textit{nonlocal}. This is an elliptical
equation, and its solution essentially depends on the (usually unphysical)
boundary conditions. A small perturbation or numerical error at the
boundary may cause a global perturbation in the full simulation domain.
In contrast, the Maxwell equations (\ref{Ampere}), (\ref{Faraday})
are local. Any signal can propagate no faster than the vacuum speed
of light, and we apply here the Minkovsky diagram, Fig.~\ref{fig:Minkovski}.
As the central event $(t_{n},x_{i},y_{j},z_{k})$, we choose some
grid cell with the indices $(n,i,j,k)$ so that $t_{n}=n*\tau$, $x_{i}=i*\Delta x$,
$y_{j}=j*\Delta y$, $z_{k}=k*\Delta z$. Here we denote $\tau,\Delta x,\Delta y,\Delta z$
the numerical steps on time, $X-$, $Y-$ and $Z-$axises. The light
cone separates the full 4-dimensional space into the regions of the
``absolute past'', the ``absolute future'', and the region of
``absolutely distant'' events. Only the events taking place in the
``absolute past'' may stay in a casual connection with the central
event. Thus, fields at the grid position $(t_{n},x_{i},y_{j},z_{k})$
are influenced by the events happening at the time moment $t_{n-1}$
at the grid cells located within the circle $c\tau$ around the original
cell. If we use an explicit numerical scheme, then the time step is
limited through the Courant condition, $c\tau<\min(\Delta x,\Delta y,\Delta z)$,
and only the first neighboring cells are involved. A numerical scheme
that does have this physical property is called \textit{local}. In
this sense, any numerical scheme that involves the solution of an
elliptical equation like the Poisson one (\ref{Poisson_phi}), is
\textit{non-local}.

It appears, that the key issue for the development of a local numerical
scheme is the method of \textit{current deposition} on the grid during
the particle motion. Let us consider the cubic FPFE (numerical particle)
on a grid. We suppose, that the particle and the grid elementary volume
(the cell volume) $V_{c}=\Delta x\Delta y\Delta z$ are \textit{identical}.
So that the particle size length $\Delta x_{\alpha}$ is also the
grid step along $\alpha-$axis, $\alpha=x,y,z$. We mark further the
grid cells with the indexes $i,j,k$ along the axises $x,y,z$.

\begin{figure}
\centerline{\includegraphics[clip,width=0.9\textwidth]{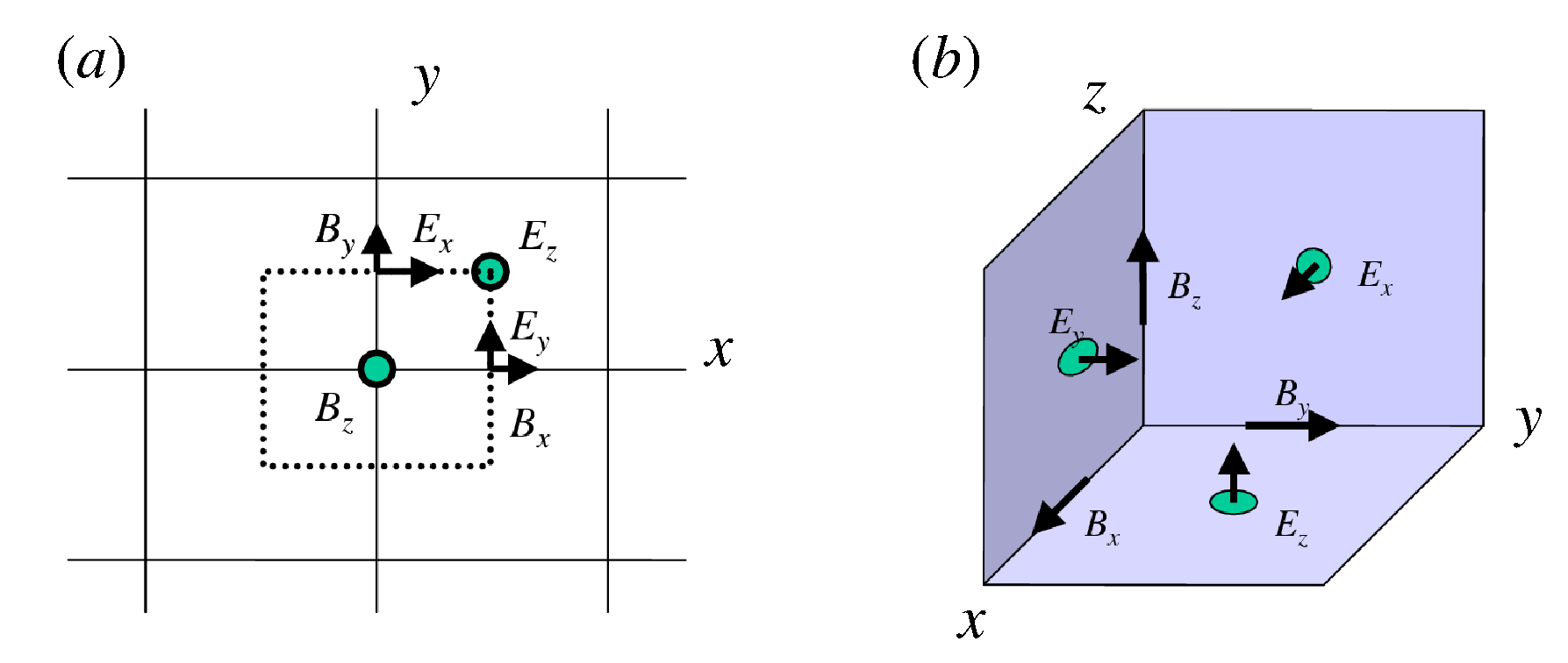}} \protect\caption{Yee lattice.}

\label{fig:YeeF} 
\end{figure}


In the discussion of multi-dimensional PIC codes, we normally use
the staggered or Yee lattice (grid), as illustrated in Fig.~\ref{fig:YeeF}.
We define the charge density on the grid at the centers of the cells,
$\rho_{i+1/2,j+1/2,k+1/2}$, we get:

\begin{equation}
\rho_{i+1/2,j+1/2,k+1/2}=\sum_{n}~W_{n}^{\rho}~S^{\rho}({\bf x_{i+1/2,j+1/2,k+1/2}-x}_{n}).\label{rho_n}
\end{equation}

\noindent The charge density interpolation weight and form of the
particle are

\begin{eqnarray}
S^{\rho}({\bf x}) & = & S_{x}^{\rho}(x)S_{y}^{\rho}(y)S_{z}^{\rho}(z),\nonumber \\
S_{j}^{\rho}({\bf x_{j}}) & = & 1-2\frac{|x_{j}|}{\Delta_{j}}\label{rho_1}\\
 &  & |x_{j}|<0.5\Delta_{j}.\nonumber 
\end{eqnarray}

\noindent The scheme (\ref{rho_1}) is the ``volume'' (or ``area'')
weighting. It actually assigns the part of the particle residing
in a cell to the cell's center.

Now, if the particle moves, it generates current. How should one interpolate
the current to the grid cells? One may try to use a straightforward
interpolation, say, ${\bf J}=\sum_{n}{\bf V_{n}}S_{n}^{\rho}$, or
others discussed in detail in Birdsall~\&~Langdon book \cite{Langdon}.
Quite naturally, these voluntary interpolations do not satisfy the
continuity equation, i.e., the current flux through a cell boundaries
defined in this way does not represent the actual charge change in
the cell. The further consequence of this procedure appears when we
integrate in time the Ampere law (\ref{Ampere}). The obtained electric
field just does not satisfy the Gauss law (\ref{Poisson}).

One of the possible solutions of this inconsistency is to \textit{correct}
the obtained electric field \cite{Langdon}. Suppose, we have advanced
the electric field ${\bf E'}$ according to (\ref{Ampere}), and we
got difficulties with the Gauss law: $\nabla\cdot{\bf E'}\ne4\pi\rho$.
We may try to correct the electric field. We introduce a potential
$\delta\phi$, such that

\begin{equation}
\nabla^{2}\delta\phi=-4\pi\rho+{\bf E'},\label{correct_phi}
\end{equation}

\noindent and construct the corrected electric field

\begin{equation}
{\bf E}={\bf E}-\nabla\delta\phi,\label{correct_E}
\end{equation}

\noindent which does satisfy the Gauss law:

\begin{equation}
\nabla{\bf E}=4\pi\rho.\label{Gauss1}
\end{equation}

\noindent Unfortunately, this correction requires a solution of the
nonlocal elliptical problem (\ref{correct_phi}).

\begin{figure}
\label{fig:pmove3D} \centerline{\includegraphics[width=11cm]{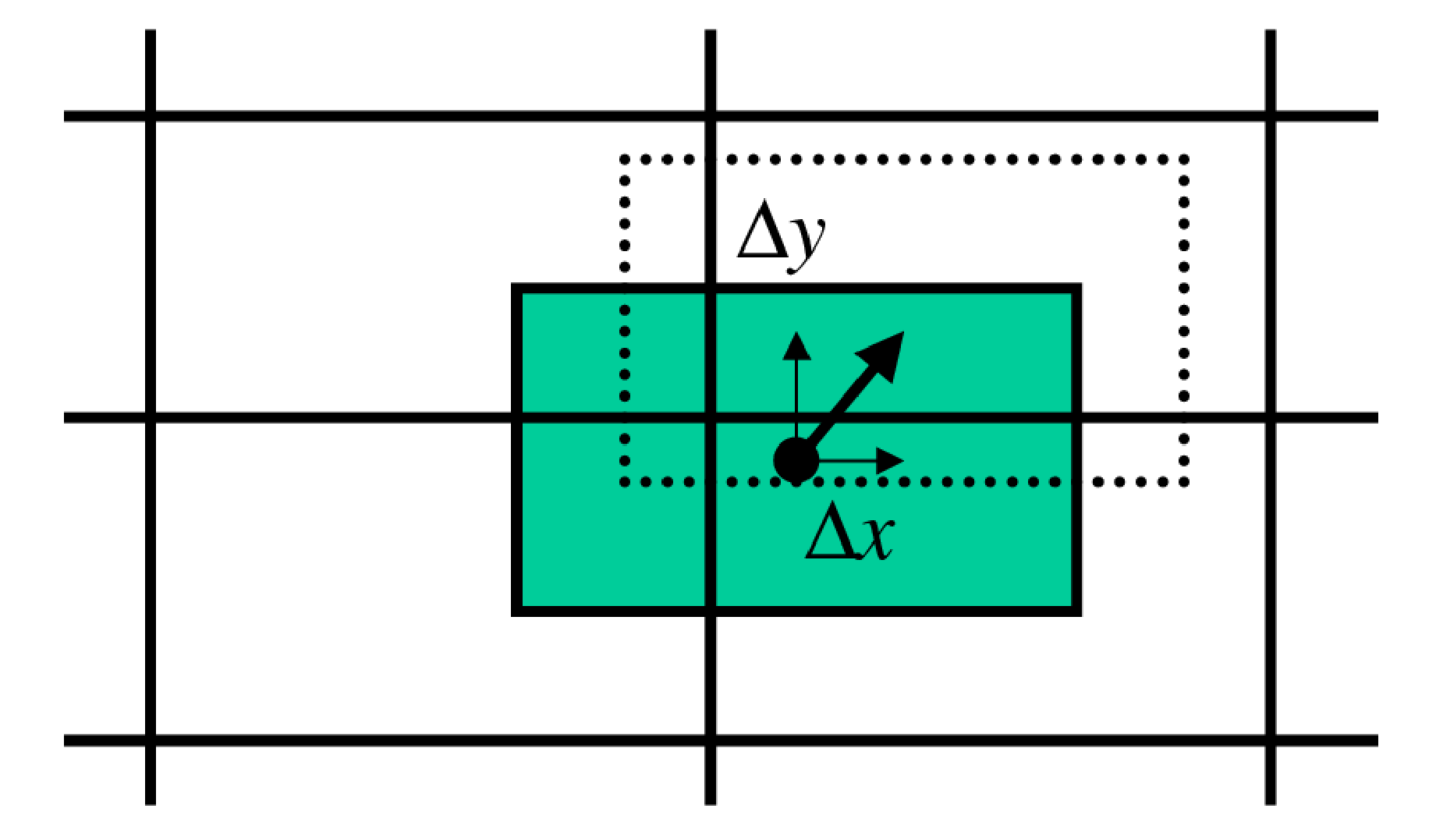}}
\protect\caption{Tracing of particle trajectory}
\end{figure}


It appears, however, that the currents can be defined in a self-consistent
way \cite{Buneman}. To do so, one has to follow the particle trajectory
in the very detail, and keep recording of how much charge has passed
through each of the cell's boundaries. Fig.~\ref{fig:pmove3D} illustrates
the idea. Let us take a particle with a center located within the
grid elementary volume $V_{c}$ centered at the grid vertex $(i,j,k)$.
The particle initial position is at $(x_{0},y_{0},z_{0})$, and after
one time step the particle moves to the new position $(x_{1},y_{1},z_{1})$.
For the first time we suppose, that the new position is still within
the elementary volume. The particle displacement is $(\delta x,\delta y,\delta z)$.
The particle generates the instantaneous current density ${\bf j}={\bf V}W^{\rho}S^{\rho}$,
and the current fluency ${\bf J}$, i.e., the charge that has crossed
some surface $\Omega$ during the time step $\tau$ is given by

\begin{equation}
{\bf J}=\int_{\Omega}d{\bf \Omega}\int_{0}^{\tau}{\bf V}W^{\rho}S^{\rho}~dt=\int_{\Omega}d\Omega\int_{{\bf x_{0}}}^{{\bf x_{1}}}W^{\rho}S^{\rho}~d{\bf x}.\label{q_passed}
\end{equation}

\noindent Thus, we have to integrate along the particle trajectory.
If, as usual \cite{Langdon}, we are using the second order finite
difference scheme to advance the particle position, then the particle
moves along a straight line during one time step. Assuming this, it
is easy to calculate that if the particle remained within the elementary
volume, it has induced the following currents on the grid:

\begin{eqnarray}
J_{i,j+1/2,k+1/2}^{x} & = & \delta xW^{\rho}(a_{y}a_{z}+b_{yz}),\nonumber \\
J_{i+1/2,j,k+1/2}^{y} & = & \delta yW^{\rho}(a_{z}a_{x}+b_{zx}),\nonumber \\
J_{i+1/2,j+1/2,k}^{z} & = & \delta zW^{\rho}(a_{x}a_{y}+b_{xy}),\nonumber \\
\nonumber \\
J_{i,j-1/2,k+1/2}^{x} & = & \delta xW^{\rho}[(1-a_{y})a_{z}-b_{yz}],\nonumber \\
J_{i,j+1/2,k-1/2}^{x} & = & \delta xW^{\rho}[a_{y}(1-a_{z})-b_{yz}],\nonumber \\
J_{i,j-1/2,k-1/2}^{x} & = & \delta xW^{\rho}[(1-a_{y})(1-a_{z})+b_{yz}],\nonumber \\
\nonumber \\
J_{i+1/2,j,k-1/2}^{y} & = & \delta yW^{\rho}[(1-a_{z})a_{x}-b_{zx}],\nonumber \\
J_{i-1/2,j,k+1/2}^{y} & = & \delta yW^{\rho}[a_{z}(1-a_{x})-b_{zx}],\nonumber \\
J_{i-1/2,j,k-1/2}^{y} & = & \delta yW^{\rho}[(1-a_{z})(1-a_{x})+b_{zx}],\nonumber \\
\nonumber \\
J_{i-1/2,j+1/2,k}^{z} & = & \delta zW^{\rho}[(1-a_{x})a_{y}-b_{xy}],\nonumber \\
J_{i+1/2,j-1/2,k}^{z} & = & \delta zW^{\rho}[a_{x}(1-a_{y})-b_{xy}],\nonumber \\
J_{i-1/2,j-1/2,k}^{z} & = & \delta zW^{\rho}[(1-a_{x})(1-a_{y})+b_{xy}],\label{addcurr}
\end{eqnarray}

\noindent where

\begin{eqnarray}
a_{\alpha} & = & 1-2\frac{|x_{\alpha}+0.5\Delta_{\alpha}|}{\Delta_{\alpha}},\label{addcurr_weight}\\
b_{\alpha\beta} & = & \frac{1}{12}\delta x_{\alpha}\delta x_{\beta};\nonumber \\
 &  & \alpha,\beta=\{x,y,z\}.\nonumber \\
\end{eqnarray}

\noindent If the particle leaves the elementary volume where it was
residing initially, then the full displacement must be split into
several ``elementary'' motions. During each of these elementary
motions the particle remains inside an elementary volume surrounding
the corresponding vertex of the grid. This ``bookkeeping'' of the
particle motion does require some programming effort. However, it
appears to be very important to implement it.

The electric field is then advanced in time according to:

\begin{equation}
{\bf E}^{n+1}-{\bf E}^{n}=c\tau\nabla{\bf \times B}^{n+1/2}-4\pi{\bf J}^{n+1/2},\label{E_advance}
\end{equation}

\noindent with the particular components of electric field ${\bf E}$
defined at the same positions on the grid as the current ${\bf J}$,
and $\hat{\nabla}{\bf \times}$ be the finite-difference version of
the curl operator.

One may think about an alternative approach which is somewhat easier
from programming point of view. Why don't we replace the actual, straight,
motion of the particle during one time step by \textit{averaging}
of all possible \textit{rectangular} paths along the grid axises which
connect the initial and the final positions of the particle? This
approach has been used by \textit{Morse \& Nielson}~(1971) \cite{Morse}.
However, this fake integration has lead to an unacceptably rapid growth
of \textit{electromagnetic} noise in their code. The author also finds
that even small deviations from the accurate current deposition (\ref{addcurr})
immediately result in noise boosting, even if the deviated scheme
is still charge conserving.

Thus, the scheme (\ref{addcurr}) is \textit{the} method to avoid
the solution of elliptical equations and still to satisfy the Gauss
law numerically. In other words, we rigorously enforce the \textit{detailed},
i.e., down to each grid cell, charge conservation and correct continuity
equation.

\section{Energy conservation}

In the previous section we have discussed how to develop an electromagnetic
Particle-in-Cell code which is rigorously charge conserving. Another
important conservation law one would like to enforce, is the total
energy conservation. Indeed, it is well known, that one of the worst
plagues spoiling the standard PIC codes is the effect of numerical
heating. The numerical ``temperature'' (rather, the chaotic energy
per numerical particle) is known to grow exponentially until the effective
Debye length becomes comparable with the grid size. Then, the exponential
growth goes over in a more moderate linear heating. This is the effect
of ``aliasing'': the inconsistent interpolation of the fields defined
on the grid to the particle position.

As we already hinted in the Introduction, there is no \textit{fundamental}
reason for the numerical heating, if we adhere to the paradigm of
Finite Phase Fluid Elements. Now we proceed to design the \textit{energy
conserving electromagnetic code}.

We start with the exact analytical equation for the full energy of
the system:

\begin{equation}
H=\sum_{n}m_{n}c^{2}(\gamma-1)~+~\frac{1}{8\pi}\int_{V}(E^{2}+B^{2})dV,\label{energy}
\end{equation}

\noindent where $m_{n}$ is the particle's mass, $\gamma=\sqrt{1+(p/m_{n}c)^{2}}$
is the relativistic $\gamma-$factor, and the integration is taken
over the full volume $V$.

Now we split the electric field into longitudinal ${\bf E_{\parallel}}$
and transverse ${\bf E_{\perp}}$ parts, so that

\begin{eqnarray}
\nabla\cdot{\bf E_{\perp}} & = & 0\nonumber \\
\nabla{\bf \times E_{\parallel}} & = & 0\label{splitE}
\end{eqnarray}

\noindent Then, we introduce a potential $\phi$ such that ${\bf E_{\parallel}}=-\nabla\phi$.
One shows easily that

\begin{equation}
\int_{V}{\bf E_{\perp}E_{\parallel}}dV=-\int_{\Omega}\phi{\bf E_{\perp}}d{\bf \Omega}=0\label{crossE}
\end{equation}

\noindent for an infinite or periodic volume. Here $\Omega$ is a
surface surrounding the volume. As a consequence, we write the energy
of our system as

\begin{equation}
H=\sum_{n}m_{n}c^{2}(\gamma-1)~+~\frac{1}{8\pi}\int_{V}(E_{\parallel}^{2}+E_{\perp}^{2}+B^{2})dV=H_{kin}+H_{s}+H_{EM}+H_{B},\label{H}
\end{equation}

\noindent where

\begin{equation}
H_{kin}=\sum_{p}m_{p}c^{2}(\gamma-1),\label{H_p}
\end{equation}

\noindent is the kinetic energy of the particles,

\begin{equation}
H_{s}=\frac{1}{8\pi}\int_{V}E_{\parallel}^{2}dV,\label{H_s}
\end{equation}

\noindent is the electrostatic part of the electric field energy,

\begin{equation}
H_{EM}=\frac{1}{8\pi}\int_{V}E_{\perp}^{2}dV,\label{H_=00007BEM=00007D}
\end{equation}

\noindent is the electromagnetic part of the electric field energy,

\begin{equation}
H_{B}=\frac{1}{8\pi}\int_{V}B^{2}dV..\label{H_B}
\end{equation}

\noindent is the magnetic field energy.

The energy (\ref{H}) has been written for continuous fields and individual
particles. A straightforward analog, however, may be written for a
finite-difference numerical scheme.

We are using the staggered grid (Yee lattice) and have fixed the current
interpolation to the grid (\ref{addcurr}). Now we have to define
the \textit{force interpolation} to the actual particle position in
such a way, that the resulting numerical scheme conserves the Hamiltonian
(\ref{H}).

The most dangerous for the numerical heating is the \textit{electrostatic}
part of the code. It is the electrostatic plasma waves that are responsible
for the Debye shielding. Also, the ${\bf v\times B}$ part of the
Lorentz force acting on the particle conserves energy automatically,
as well as the ${\bf B-}$field advance according to the Faraday law
(\ref{Faraday}). Thus, for the moment, we forget about the magnetic
field and the magnetic energy part. We enforce conservation of $H_{E}=H_{kin}+H_{s}+H_{EM}$.
The numerical scheme will conserve the energy if it is derived from
equations in the canonical form:

\begin{eqnarray}
\frac{d {\bf p}_{p}}{d t} & = & -\partial_{{\bf x_{p}}}H_{E},\label{p_can}\\
\frac{ d{\bf x}_{p}} {d t}& = & \partial_{{\bf p_{p}}}H_{E},\label{x_can}
\end{eqnarray}

\noindent where the index $p$ runs through all particles.

We rewrite Eqs.~(\ref{p_can})-(\ref{x_can}) more explicitly:

\begin{eqnarray}
\frac{ d{\bf p}_{p}}{d t} & = & -\frac{1}{4\pi}\int_{V}{\bf E}\cdot\partial_{{\bf x_{p}}}{\bf E}dV,\label{p_can_e}\\
\frac{ d{\bf x}_{p}} {d t} & = & \frac{{\bf p}_{p}}{\gamma_{p}}={\bf V}_{p}.\label{x_can_e}
\end{eqnarray}

\noindent To deal with the Eq.~(\ref{p_can_e}), one has to refer
to the equation on the electric field advance in time (\ref{E_advance}).
To get the correct expression for $\partial_{{\bf x_{p}}}{\bf E}$,
let us displace the particle $p$ by a small distance ${\bf \delta x}$.
This displacement generates a current ${\bf \delta J}$ on the adjacent
grid positions in accordance with (\ref{addcurr}). The resulting
change ${\bf \delta E}$ of the electric field is

\begin{equation}
{\bf \delta E}=-4\pi{\bf \delta J}\label{dE}
\end{equation}

\noindent at the same grid positions. Thus, we may rewrite the first
canonical Eq.~(\ref{p_can_e}) as:

\begin{equation}
\frac{ d{\bf p}_{p}}{d t}=\int_{V}{\bf E}\cdot\partial_{{\bf x_{p}}}{\bf \delta J}dV.\label{p_can_j}
\end{equation}

\noindent The expression (\ref{p_can_j}) has a very simple and clear
meaning. To make the Particle-in-Cell code energy conserving, one
has to employ the same scheme for the electric field interpolation
to the particle position as for the current deposition. Thus, the
energy conserving interpolation scheme for the $E-$field is:

\begin{eqnarray}
E_{x}^{p} & = & W^{\rho}[E_{i,j+1/2,k+1/2}^{x}a_{y}a_{z}+\nonumber \\
 &  & E_{i,j-1/2,k+1/2}^{x}(1-a_{y})a_{z}+\nonumber \\
 &  & E_{i,j+1/2,k-1/2}^{x}a_{y}(1-a_{z})+\nonumber \\
 &  & E_{i,j-1/2,k-1/2}^{x}(1-a_{y})(1-a_{z})],\nonumber \\
E_{y}^{p} & = & W^{\rho}[E_{i+1/2,j,k+1/2}^{y}a_{x}a_{z}+\nonumber \\
 &  & E_{i-1/2,j,k+1/2}^{y}(1-a_{x})a_{z}+\nonumber \\
 &  & E_{i+1/2,j,k-1/2}^{y}a_{x}(1-a_{z})+\nonumber \\
 &  & E_{i-1/2,j,k-1/2}^{y}(1-a_{x})(1-a_{z})],\nonumber \\
E_{z}^{p} & = & W^{\rho}[E_{i+1/2,j+1/2,k}^{z}a_{y}a_{x}+\nonumber \\
 &  & E_{i+1/2,j-1/2,k}^{z}(1-a_{y})a_{x}+\nonumber \\
 &  & E_{i-1/2,j+1/2,k}^{z}a_{y}(1-a_{x})+\nonumber \\
 &  & E_{i-1/2,j-1/2,k}^{z}(1-a_{y})(1-a_{x})],\label{E_inter}
\end{eqnarray}

\noindent It is important, that the electric field is taken\textit{at
the present particle position} and not averaged along the trajectory.
Also, the higher-order corrections $b_{\alpha}$, which we have introduced
for the current depositions, are absent. This is because Eq.~(\ref{p_can_j})
gives the analytical expression for the infinitesimal particle displacements,
i.e., has to be considered in the limit $|{\bf \delta x}|\rightarrow0$.

\subsection{Particle push}

For the particle advance in time one might then use the Boris scheme
\cite{Langdon} with the electric field ${\bf E}$ interpolated according
to (\ref{E_inter}) and the magnetic field ${\bf B}$ that can be
interpolated using a different scheme, as discussed later. The Boris
scheme is

\begin{equation}
\frac{{\bf p}_{1}-{\bf p}_{0}}{\tau}=e\left({\bf E}+\frac{1}{c}\frac{{\bf p}_{1}+{\bf p}_{0}}{2\gamma_{1/2}}{\bf ~\times~B}\right).\label{Boris}
\end{equation}

\noindent where ${\bf p}_{0}$ and ${\bf p}_{1}$ are the initial
and the final particle momenta, $\gamma_{1/2}$ is the $\gamma-$factor
taken at the middle of the time step. This scheme is time-reversible,
and \textit{semi-}implicit. It can be analytically resolved for the
final momentum ${\bf p}_{1}$ \cite{Langdon}:

\begin{eqnarray}
{\bf p}^{n+1/2}={\bf p}^{-}-e{\bf E}\frac{\tau}{2}\label{eq:half step}\\
{\bf p}^{n-1/2}={\bf p}^{+}+e{\bf E}\frac{\tau}{2}\\
\frac{{\bf p}^{+}-{\bf p}^{-}}{\tau}=\frac{q}{2\gamma mc}{\bf p}^{+}+{\bf p}^{-}\times{\bf B}\\
{\bf p}'={\bf p}^{-}+\frac{q\tau}{2\gamma}{\bf p}^{-}\times{\bf B}\\
{\bf p}^{+}={\bf p}^{-}+\frac{2}{1+\left(\frac{q\tau B}{2\gamma}\right)^{2}}{\bf p}'\times{\bf B}.
\end{eqnarray}

The $\gamma-$factor should be calculated after the step \eqref{eq:half step}.
However, as the magnetic field rotates the particle momentum, the
Boris scheme is not exactly symmetric. An alternative scheme has been
proposed recently \cite{Vaypusher}. We derive this alternative scheme
below.

The equation of motion is discretized as: 
\begin{equation}
\frac{\mathbf{p}-\mathbf{p}_{0}}{\tau}=q\mathbf{E}+\frac{q}{2}\left(\frac{\mathbf{p}}{\gamma}+\frac{\mathbf{p}_{0}}{\gamma_{0}}\right)\times\mathbf{B}\label{eq:basic}
\end{equation}
Here, ${\bf p_{0}}$ is the initial particle momentum and ${\bf p}$
is the particle momentum after the push with the corresponding $\gamma-$factors.
We can rewrite this equation in the form

\begin{equation}
\mathbf{p}=\mathbf{a}+\frac{\mathbf{p}}{\gamma}\times\mathbf{b}\label{eq:simple}
\end{equation}
where 
\begin{equation}
\mathbf{a}=\mathbf{p}_{0}+q\tau\mathbf{E}+\frac{q\tau}{2}\frac{\mathbf{p}_{0}}{\gamma_{0}}\times\mathbf{B}\label{eq:a}
\end{equation}
and

\begin{equation}
\mathbf{b}=\frac{q\tau}{2}\mathbf{B}\label{eq:b}
\end{equation}
We rewrite \eqref{eq:simple} as
\begin{equation}
\gamma\mathbf{p}=\gamma\mathbf{a}+\mathbf{p}\times\mathbf{b}\label{eq:simple-1}
\end{equation}
Scalar multiply \eqref{eq:simple-1} with $\mathbf{p}$ gives

\begin{equation}
p^{2}=\mathbf{a}\cdot\mathbf{p}\label{eq:ap}
\end{equation}
Scalar multiply \eqref{eq:simple-1} with $\mathbf{b}$ gives

\begin{equation}
\mathbf{b}\cdot\mathbf{p}=\mathbf{a}\cdot\mathbf{b}\label{eq:bp}
\end{equation}
Scalar multiply \eqref{eq:simple-1} with $\mathbf{a}$ gives

\begin{equation}
\gamma\mathbf{a}\cdot\mathbf{p}-\gamma a^{2}=\mathbf{a}\cdot\left(\mathbf{p}\times\mathbf{b}\right)=\mathbf{p}\cdot\left(\mathbf{b}\times\mathbf{a}\right)=\mathbf{b}\cdot\left(\mathbf{a}\times\mathbf{p}\right)\label{aa}
\end{equation}
Vector multiply $\mathbf{a}$ with \eqref{eq:simple-1} gives

\begin{equation}
\gamma\mathbf{a}\times\mathbf{p}=\mathbf{p}\left(\mathbf{a}\cdot\mathbf{b}\right)-\mathbf{b}\left(\mathbf{a}\cdot\mathbf{p}\right)\label{aXp-1}
\end{equation}
Scalar multiply \eqref{eq:simple-1} with $\mathbf{a}$ gives

\begin{equation}
\gamma\left(p^{2}-a^{2}\right)=\frac{\mathbf{b}}{\gamma}\left[\mathbf{p}\left(\mathbf{a}\cdot\mathbf{b}\right)-\mathbf{b}\left(\mathbf{a}\cdot\mathbf{p}\right)\right]=\frac{\left[\left(\mathbf{a}\cdot\mathbf{b}\right)^{2}-b^{2}p^{2}\right]}{\gamma}\label{aa-1}
\end{equation}
or
\begin{equation}
\gamma^{2}\left(\gamma^{2}-1-a^{2}\right)=\left(\mathbf{a}\cdot\mathbf{b}\right)^{2}-b^{2}\left(\gamma^{2}-1\right)\label{gamma2}
\end{equation}
This leads to the quadratic equation
\begin{equation}
\gamma^{4}+\gamma^{2}\left(b^{2}-1-a^{2}\right)-b^{2}-\left(\mathbf{a}\cdot\mathbf{b}\right)^{2}=0\label{gamma2-1}
\end{equation}
with the solution

\begin{equation}
\gamma^{2}=\frac{1+a^{2}-b^{2}}{2}+\sqrt{\left(\frac{1+a^{2}-b^{2}}{2}\right)^{2}+b^{2}+\left(\mathbf{a}\cdot b\right)^{2}}\label{gamma2-2}
\end{equation}
Now we have to find the particle momentum after the push$\mathbf{p}$.
For this, we vector multiply $\mathbf{b}$ with \eqref{eq:simple-1}:

\begin{equation}
\gamma\left(\mathbf{\mathbf{b}}\times\mathbf{p}-\mathbf{b}\times\mathbf{a}\right)=\mathbf{p}\cdot\mathbf{b}-\mathbf{b}\left(\mathbf{p}\cdot\mathbf{b}\right)=\mathbf{p}b^{2}-\mathbf{b}\left(\mathbf{a}\cdot\mathbf{b}\right)\label{bXp}
\end{equation}
Using \eqref{eq:simple-1}, we find

\begin{equation}
\gamma^{2}\mathbf{a}-\gamma^{2}\mathbf{p}-\gamma\mathbf{b}\times\mathbf{a}=\mathbf{p}b^{2}-\mathbf{b}\left(\mathbf{a}\cdot\mathbf{b}\right)\label{bXp-1}
\end{equation}
Solving for $\mathbf{p}$, we obtain

\begin{equation}
\mathbf{p}=\frac{\gamma^{2}\mathbf{a}+\gamma\mathbf{a}\times\mathbf{b}+\mathbf{b}\left(\mathbf{a}\cdot\mathbf{b}\right)}{\gamma^{2}+b^{2}}\label{new p}
\end{equation}
This pusher is fully implicit and does not require splitting of the
Lorentz operator in electric field push and magnetic field rotation.

\subsection{Energy conservation tests}

The interpolation scheme (\ref{addcurr}),(\ref{E_inter}) conserves
the energy \textit{exactly} for time steps, which are small enough
and the particle does not leave the original grid cell. If, however,
motion of the particle becomes highly relativistic, the system does
exhibit a slow energy growth. Notwithstanding this small drawback,
we have solved one of the major problems in PIC codes. We may now
simulate \textit{cold} plasma. As the ``cold'' usually means non-relativistic
``temperatures'', the energy is conserved, and the numerical heating
is absent. If we do have a hot plasma, with temperatures close to
relativistic ones, the approach of \textit{stochastic} sampling of
the phase space becomes valid. Fortunately, the Debye length of such
plasma is many grid cells anyway, and the numerical heating is not
an issue.

The scheme for the electric field interpolation to the particle position
(\ref{E_inter}) is identical to the energy-conserving one used in
electrostatic codes with charge-potential, $(\rho-\phi)$, formalism
\cite{Langdon,Hockney}. However, as we will see further, there is
a significant difference between these electrostatic codes and the
fields-current, $({\bf E,J})$ formalism used in our electromagnetic
simulations.

\begin{figure}
\centerline{\includegraphics[width=0.9\textwidth]{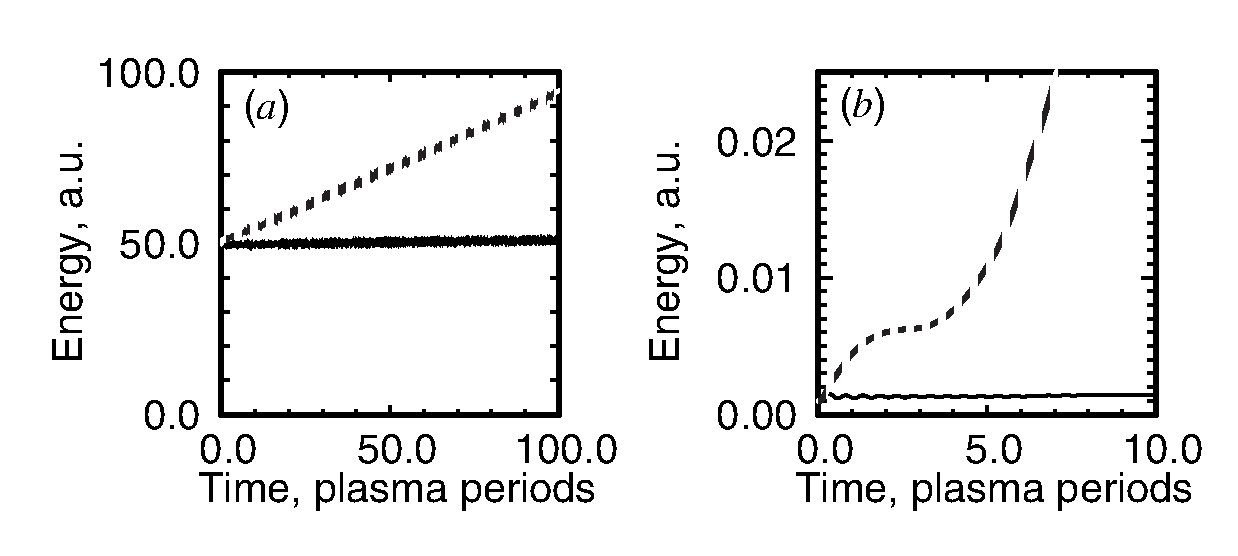}}
 \protect\caption{ Energy conservation in VLPL code due to EC algorithm (solid lines)
and numerical heating in standard MC algorithm (dashed lines) when
(a) the Debye length $D=0.5\Delta x$, warm plasma and (b) $D=5\cdot10^{-3}\Delta x$,
cold plasma. Energy change in VLPL code is within 2\% after 100 plasma
periods. }

\label{econs} 
\end{figure}


The Fig.~\ref{econs} shows the total energy evolution in an isolated
system of particles for an energy conservative (solid lines) and a
``momentum conservative'' (MC) (\cite{Langdon}) (dashed lines)
algorithms for the two cases: (a) Debye length is $D=0.5h$ and (b)
$D=5\times10^{-3}h$. Although the energy conservation for the EC
algorithm is not exact, it is much better than for the MC case. The
actual energy change is only 2\% over 100 plasma oscillations for
the EC algorithm. This property makes it possible to simulate a cold
plasma.

\section{Momentum and current conservation}

It is known, that the energy-conserving electrostatic PIC codes do
not conserve momentum \cite{Langdon}. Indeed, it is easy to show
that, strictly speaking, the electric field interpolation to the particle
position in the form of (\ref{E_inter}) does not conserve the total
momentum

\begin{equation}
\frac{d {\bf P_{total}}}{dt}=\sum_{p}q_{p}E_{p}\ne0,\label{p_inconserve}
\end{equation}

\noindent if the particles cross cell boundaries during their motion.

Now the question is: how detrimental is this lack of momentum conservation
for the PIC code?

The momentum nonconservation in electrostatic PIC codes using the $(\rho-\phi)$
formalism leads to impressive consequences \cite{Langdon}. If one
starts the simulation with electrons drifting with respect to the
resting ions, the electrons experience some average \textit{drag force}
from the grid. As a result, after a few plasma periods, an initially
regular electron drift becomes chaotic and the simulation ends with disordered
hot electrons without any net motion with respect to the ions. Although
the final energy of the system is preserved, and remains the same
as the initial kinetic energy of the drifting electrons, the momentum
conservation failure is spectacular.

However, this spectacular example of momentum inconservation in the
one-dimensional electrostatic PIC code is rather an artefact of the
$(\rho-\phi)$ formalism. Moreover, even the inital ``equilibrium''
of electrons drifting with respect to ions is an artefact itself.
Indeed, the original Maxwell equations (\ref{Ampere})-(\ref{Faraday})
simply \textit{do not allow for freely drifting electrons in the one-dimensional
geometry!} This drift would correspond to a constant current, which
results in a fast build-up of the longitudinal electric field. As
a consequence, electrons must oscillate around their initial positions
at the local plasma frequency. Apparently, this contradiction with
the Maxwell equations remained unmentioned in the $(\rho-\phi)$ formulation
of that electrostatic code.

In the more realistic $({\bf E-J})$ formalism of electromagnetic
codes, the ions have to drift \textit{together} with the electrons,
unless the forward electron current is compensated by some artificial
``return'' current, e.g., the longitudinal part of ${\bf \nabla\times B}$,
which evidently does not exist in the 1D geometry. Anyway, a code
using the $({\bf E-J})$ conserves the net current per definition:

\begin{equation}
<{\bf J}>={\bf J_{0}}={\bf Const},\label{j_cons}
\end{equation}

\noindent where the averaging is made in time over the local plasma
frequency. Indeed, the only possible deviations in the current are
due to the longitudinal part of electric field, i.e., the \textit{charge
displacement} current.

The current conservation (\ref{j_cons}) is the very important property,
which may \textit{compensate} for the absence of a detailed momentum
conservation. As an example, let us consider the total electron current
flowing in the simulation:

\begin{equation}
{\bf J}_{e}=\sum_{p}eW_{p}{\bf V}_{p},\label{Je}
\end{equation}

\noindent where $e$ is the electron charge, and $W_{p}$ is the ``weight''
of the numerical particle $p$. When averaged over plasma period,
the current (\ref{Je}) is conserved by our code. Now, we may write
the electron momentum in the non-relativistic case as

\begin{equation}
{\bf P}_{e}=\sum_{p}m_{e}W_{p}{\bf V}_{p}=\frac{e}{m_{e}}{\bf J}_{e},\label{Pe}
\end{equation}

\noindent where $m_{e}$ is the electron mass.

It follows from (\ref{Pe}) that the total momentum is simply proportional
to the total current, and thus it is conserved on average. This is
a good news for the energy conserving PIC code we have designed here.
The code does conserve momentum on average in the non-relativistic
case. When the particles are moving with relativistic energies, however,
the identity (\ref{Pe}) breaks, and the momentum conservation is
not ideal again. Fortunately, the current conservation (\ref{Je})
is still a strong enough symmetry to prevent bad consequences like
those discussed in \cite{Langdon}.

\section{Maxwell solver: Numerical Dispersion Free scheme}

Earlier, we have discussed how to push particles and collect currents
on the grid. Now, we are discussing the finite-difference solver for
the time-dependent Maxwell Eqs.~(\ref{Faraday})-(\ref{Ampere}).

The standard approach to propagate the fields on Yee lattice, see
Fig.~\ref{fig:YeeF} is to write the centered conservative scheme
\cite{Langdon}. Let us first consider the two-dimensional geometry
for the sake of simplicity. In the 2D $(X-Y)$ geometry, one may distinguish
two kinds of polarizations, $s-$polarizations with $E_{z},B_{x},B_{y}$
fields, and $p-$polarization with $E_{x},E_{y},B_{z}$ fields. It
can be shown, that if the initial condition contains $p-$polarized
fields and $J_{x},J_{y}$ currents only, the $s-$polarized fields
are not excited at all \cite{Langdon}. Thus, we may take the $p-$polarization
as an example. The standard 2D scheme for the $p-$polarization is

\begin{eqnarray}
{B_{z}}_{i,j}^{n+1/2}-{B_{z}}_{i,j}^{n-1/2} & = & \frac{c\tau}{\Delta y}({E_{x}}_{i,j+1/2}^{n}-{E_{x}}_{i,j-1/2}^{n})\nonumber \\
 & - & \frac{c\tau}{\Delta x}({E_{y}}_{i+1/2,j}^{n}-{E_{y}}_{i-1/2,j}^{n}),\label{Bz}\\
{E_{x}}_{i,j+1/2}^{n+1}-{E_{x}}_{i,j+1/2}^{n} & = & \frac{c\tau}{\Delta y}({B_{z}}_{i,j+1}^{n+1/2}-{B_{z}}_{i,j}^{n+1/2})-4\pi\tau{j_{x}}_{i,j+1/2}^{n+1/2},\label{Ex}\\
{E_{y}}_{i+1/2,j}^{n+1}-{E_{y}}_{i+1/2,j}^{n} & = & \frac{c\tau}{\Delta x}({B_{z}}_{i+1,j}^{n+1/2}-{B_{z}}_{i,j}^{n+1/2})-4\pi\tau{j_{y}}_{i+1/2,j}^{n+1/2}.\label{Ey}
\end{eqnarray}

\noindent The scheme (\ref{Bz})-(\ref{Ey}) uses the centered expression
for the finite difference ${\bf \nabla\times}$ operators like

\begin{equation}
\left({\bf \nabla\times B}_{z}\right)_{x}=\frac{1}{\Delta y}({B_{z}}_{i,j+1}^{n+1/2}-{B_{z}}_{i,j}^{n+1/2})\label{curl}
\end{equation}

\noindent The Maxwell Eqs.~\ref{Faraday})-(\ref{Ampere}) and the
corresponding scheme (\ref{Bz})-(\ref{Ey}) are essentially linear
partial differential equation with the only nonlinear source terms
in the form of the currents ${\bf j}$. In this case, we decide about
quality of the finite difference scheme (\ref{Bz})-(\ref{Ey}) comparing
its dispersion properties with those of the Maxwell Eqs. themselves.

\begin{figure}
\centerline{\includegraphics[width=0.9 \textwidth]{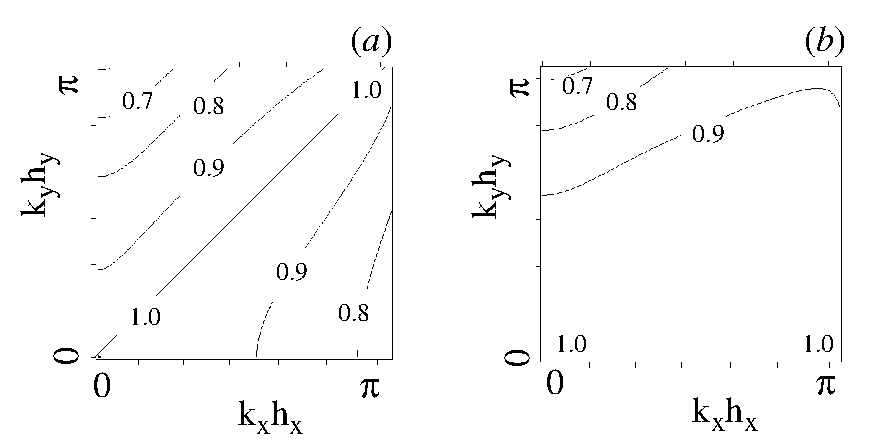}} 
 \protect\caption{Numerical phase velocity $v_{ph}/c$ corresponding (a) to the standard
scheme (\ref{DR0}) and (b) to the NDFX scheme (\ref{dispNDFX}) used
in the VLPL code. The grid cell aspect ratio $\Delta x/\Delta y=0.5$.
The scheme (\ref{DR0}) is dispersionless along the mesh diagonals,
$k_{x}=k_{y}$, while the NDFX scheme (\ref{dispNDFX}) is dispersionless
for waves running in $X-$direction when $k_{y}=0$. }

\label{fig:NDF} 
\end{figure}

According to the Maxwell Eqs., all electromagnetic waves in vacuum
run with the speed of light $c$. There is no dispersion in vacuum.
Not so for the finite differences. We Fourier-analyze the scheme (\ref{Bz})-(\ref{Ey})
by decomposing all the field in plane waves

\begin{eqnarray}
{\bf E} & = & \sum_{{\bf k}}{\bf E_{k}}\exp(-i\omega_{{\bf k}}t+i{\bf kx}),\nonumber \\
{\bf B} & = & \sum_{{\bf k}}{\bf B_{k}}\exp(-i\omega_{{\bf k}}t+i{\bf kx}),\label{modes}
\end{eqnarray}

\noindent where ${\bf k}$ is the wave vector, $\omega_{{\bf k}}$
the corresponding frequency, and ${\bf E_{k},B_{k}}$ amplitudes of
the Fourier-harmonics. For the continuum Maxwell Eqs. we have the
simple dispersion relation

\begin{equation}
\omega_{{\bf k}}=c|{\bf k}|,
\end{equation}

\noindent while the discretization in the finite differences (\ref{Bz})-(\ref{Ey})
introduces the numerical dispersion

\begin{equation}
\frac{1}{c^{2}\tau^{2}}\sin^{2}\frac{\omega_{{\bf k}}\tau}{2}=\frac{1}{\Delta x^{2}}\sin^{2}\frac{k_{x}\Delta x}{2}+\frac{1}{\Delta y^{2}}\sin^{2}\frac{k_{y}\Delta y}{2},\label{DR0}
\end{equation}

\noindent where $\Delta x$, $\Delta y$ are the grid steps, $\tau$
is the time step. The time step is limited by the Courant stability
condition: $c^{2}\tau^{2}\le\Delta x^{2}\Delta y^{2}/(\Delta x^{2}+\Delta y^{2})$.
If we run the code close to this limit of stability, then only the
waves propagating along the grid diagonals are dispersion-free. The
largest numerical dispersion experience the waves running along the
grid axises. This is illustrated in Fig.~\ref{fig:NDF}a, where we
plot the phase velocity of the numerical modes, $V_{ph}=\omega_{k}/k$,
for this standard scheme. We mention that we have chosen here $\Delta y=2\Delta x$,
and this explains the evident asymmetry of the plot.

Although the numerical dispersion of the standard scheme might be
not an issue when one simulates dense, nearly critical, plasma, it
can make troubles for simulations of very underdense plasmas, as it
is important for the problem of particle acceleration \cite{Esarey}.
In the dense plasma case, the plasma dispersion is usually stronger
than the numerical one, and the problem is masked. In the low-density
plasma, however, the plasma dispersion is small. Yet, it has to be
resolved accurately, as it influences the phase velocity of the laser
pulse. As a consequence, one has to use many grid cells per laser
wavelength to obtain physically correct results. Of course, one could
send the laser along one of the grid diagonals, but this is extremely
inconvenient from the programming point of view.

We are now out for the design of a superior numerical scheme, which
does not have a numerical dispersion at all, or removes it to a large
extent. Let us return to the finite difference expression for the
curl operator (\ref{curl}). This centered operator can be rewritten
in a different way, using averages from the adjacent cells:

\begin{equation}
\left({\bf \nabla\times B}_{z}\right)_{x}\rightarrow\frac{1}{2\Delta y}({B_{z}}_{i+1,j+1}^{n+1/2}+{B_{z}}_{i-1,j+1}^{n+1/2}-{B_{z}}_{i+1,j}^{n+1/2}-{B_{z}}_{i-1,j}^{n+1/2}).\label{curl_av}
\end{equation}

\noindent On the first sight, it is unclear, what did we gain from
the averaging. It seems evident, that the averaged scheme (\ref{curl_av})
might have only worse dispersion than the simpler one (\ref{curl}).
This is only partially correct.

It appears, however, that one may use a \textit{linear combination}
of the two schemes, (\ref{curl}) and (\ref{curl_av}), in such a
way, that the dispersions of the two scheme \textit{compensate for
each other}! The resulting scheme for the $p-$polarization in the
2D geometry is

\begin{eqnarray}
{B_{z}}_{i,j}^{n+1/2} & - & {B_{z}}_{i,j}^{n-1/2}=\frac{c\tau}{\Delta y}(b_{x}({E_{x}}_{i,j+1/2}^{n}-{E_{x}}_{i,j-1/2}^{n})\nonumber \\
 & + & a_{x}({E_{x}}_{i+1,j+1/2}^{n}-{E_{x}}_{i+1,j-1/2}^{n}+{E_{x}}_{i-1,j+1/2}^{n}-{E_{x}}_{i-1,j-1/2}^{n}))\nonumber \\
 & - & \frac{c\tau}{\Delta x}(b_{y}({E_{y}}_{i+1/2,j}^{n}-{E_{y}}_{i-1/2,j}^{n})\nonumber \\
 & + & a_{y}({E_{y}}_{i+1/2,j+1}^{n}-{E_{y}}_{i-1/2,j+1}^{n}+{E_{y}}_{i-1/2,j-1}^{n}-{E_{y}}_{i-1/2,j-1}^{n})),\label{Bz1}\\
\nonumber \\
{E_{x}}_{i,j+1/2}^{n+1} & - & {E_{x}}_{i,j+1/2}^{n}=\frac{c\tau}{\Delta y}({B_{z}}_{i,j+1}^{n+1/2}-{B_{z}}_{i,j}^{n+1/2})-4\pi\tau{j_{x}}_{i,j+1/2}^{n+1/2},\label{Ex1}\\
\nonumber \\
{E_{y}}_{i+1/2,j}^{n+1} & - & {E_{y}}_{i+1/2,j}^{n}=\frac{c\tau}{\Delta x}({B_{z}}_{i+1,j}^{n+1/2}-{B_{z}}_{i,j}^{n+1/2})-4\pi\tau{j_{y}}_{i+1/2,j}^{n+1/2},\label{Ey1}
\end{eqnarray}

\noindent where the coefficients of the linear the combination of
the two schemes is

\begin{eqnarray}
a_{x} & = & a_{y}=0.125\frac{\Delta x}{\Delta y},\nonumber \\
b_{x} & = & 1-2a_{x},\nonumber \\
b_{y} & = & 1-2a_{y},\label{NDFcoeffs2d}
\end{eqnarray}

\noindent and we have supposed that $\Delta x\le\Delta y$.

The dispersion relation for numerical scheme (\ref{Bz1})-(\ref{Ey1})
is immediately found to be

\begin{eqnarray}
\frac{1}{c^{2}\tau^{2}}\sin^{2}\frac{\omega\tau}{2} & = & \frac{1}{\Delta x^{2}}\sin^{2}\frac{k_{x}\Delta x}{2}(b_{y}+2a_{y}\cos k_{y}\Delta y)\nonumber \\
 & + & \frac{1}{\Delta y^{2}}\sin^{2}\frac{k_{y}\Delta y}{2}(b_{x}+2a_{x}\cos k_{x}\Delta x).\label{dispNDFX}
\end{eqnarray}

\noindent It follows from (\ref{dispNDFX}) that the scheme is stable
even at $c\tau=\Delta x$. This is a quite unique property for an
explicit multi-dimensional finite-difference scheme. In addition,
the scheme (\ref{dispNDFX}) goes over in the usual Yee scheme in
the limit $\Delta_{x}/\Delta_{y}->0$.

When used close to the stability limit, the scheme completely removes
numerical dispersion along the $X-$axis (the laser propagation direction).
For this reason we call this scheme \textbf{NDF}: Numerical Dispersion
Free \cite{VLPL}. The phase velocity of the numerical modes for the
\textbf{NDF} scheme we plot in Fig.~\ref{fig:NDF}b. We mention,
that the region where the numerical phase velocities are close to
$c$ becomes much wider than for the standard scheme, Fig.~\ref{fig:NDF}a.

The plasma presence changes the stability condition slightly, and
the maximum $\tau$ is limited to be:

\begin{equation}
1-\frac{\tau}{\Delta x}>\frac{\omega_{p}^{2}\tau^{2}}{4},\label{taup}
\end{equation}

\noindent where $\omega_{p}=\sqrt{4\pi n_{e}e^{2}/m_{e}}$ is the
maximum plasma frequency in the simulation domain. For an underdense
plasma, however, this is an insignificant change.

The scheme (\ref{Bz1})-(\ref{Ey1}) is written for the $p-$polarization
in the 2D planar geometry. It is must be slightly modified before
it can be used in the full 3D space. The final version of the 3D \textbf{NDF}
scheme is:

\begin{eqnarray}
{B_{x}}_{i+1/2,j,k}^{n+1/2} & - & {B_{x}}_{i+1/2,j,k}^{n-1/2}=-\frac{c\tau}{\Delta y}(b_{z}({E_{z}}_{i+1/2,j+1/2,k}^{n}-{E_{z}}_{i+1/2,j-1/2,k}^{n})\nonumber \\
 & + & a_{z}({E_{z}}_{i+1/2,j+1/2,k+1}^{n}-{E_{z}}_{i+1/2,j-1/2,k+1}^{n}+\nonumber \\
 &  & {E_{z}}_{i+1/2,j+1/2,k-1}^{n}-{E_{z}}_{i+1/2,j-1/2,k-1}^{n}))\nonumber \\
 & + & \frac{c\tau}{\Delta z}(b_{y}({E_{y}}_{i+1/2,j,k+1/2}^{n}-{E_{y}}_{i+1/2,j,k-1/2}^{n})\nonumber \\
 & + & a_{y}({E_{y}}_{i+1/2,j+1,k+1/2}^{n}-{E_{y}}_{i+1/2,j+1,k-1/2}^{n}+\nonumber \\
 &  & {E_{y}}_{i+1/2,j-1,k+1/2}^{n}-{E_{y}}_{i+1/2,j-1,k-1/2}^{n})),\label{Bx3}\\
\nonumber \\
{B_{y}}_{i,j+1/2,k}^{n+1/2} & - & {B_{x}}_{i+1/2,j,k}^{n-1/2}=\frac{c\tau}{\Delta x}(b_{z}({E_{z}}_{i+1/2,j+1/2,k}^{n}-{E_{z}}_{i-1/2,j+1/2,k}^{n})\nonumber \\
 & + & a_{z}({E_{z}}_{i+1/2,j+1/2,k+1}^{n}-{E_{z}}_{i-1/2,j+1/2,k+1}^{n}+\nonumber \\
 &  & {E_{z}}_{i+1/2,j+1/2,k-1}^{n}-{E_{z}}_{i-1/2,j+1/2,k-1}^{n}))\nonumber \\
 & - & \frac{c\tau}{\Delta z}(b_{x}({E_{x}}_{i,j+1/2,k+1/2}^{n}-{E_{x}}_{i,j+1/2,k-1/2}^{n})\nonumber \\
 & + & a_{x}({E_{x}}_{i+1,j+1/2,k+1/2}^{n}-{E_{x}}_{i+1,j+1/2,k-1/2}^{n}+\nonumber \\
 &  & {E_{x}}_{i-1,j+1/2,k+1/2}^{n}-{E_{x}}_{i-1,j+1/2,k-1/2}^{n})),\label{By3}\\
\nonumber \\
{B_{z}}_{i,j,k+1/2}^{n+1/2} & - & {B_{z}}_{i,j,k+1/2}^{n-1/2}=\frac{c\tau}{\Delta y}(b_{x}({E_{x}}_{i,j+1/2,k+1/2}^{n}-{E_{x}}_{i,j-1/2,k+1/2}^{n})\nonumber \\
 & + & a_{x}({E_{x}}_{i+1,j+1/2,k+1/2}^{n}-{E_{x}}_{i+1,j-1/2,k+1/2}^{n}+\nonumber \\
 &  & {E_{x}}_{i-1,j+1/2,k+1/2}^{n}-{E_{x}}_{i-1,j-1/2,k+1/2}^{n}))\nonumber \\
 & - & \frac{c\tau}{\Delta x}(b_{y}({E_{y}}_{i+1/2,j,k+1/2}^{n}-{E_{y}}_{i-1/2,j,k+1/2}^{n})\nonumber \\
 & + & a_{y}({E_{y}}_{i+1/2,j+1,k+1/2}^{n}-{E_{y}}_{i-1/2,j+1,k+1/2}^{n}+\nonumber \\
 &  & {E_{y}}_{i-1/2,j-1,k+1/2}^{n}-{E_{y}}_{i-1/2,j-1,k+1/2}^{n})),\label{Bz3}\\
\nonumber \\
{E_{x}}_{i,j+1/2,k+1/2}^{n+1} & - & {E_{x}}_{i,j+1/2,k+1/2}^{n}=\frac{c\tau}{\Delta y}(b_{z}({B_{z}}_{i,j+1,k+1/2}^{n+1/2}-{B_{z}}_{i,j,k+1/2}^{n+1/2})\nonumber \\
 & + & a_{z}({B_{z}}_{i,j+1,k+3/2}^{n+1/2}-{B_{z}}_{i,j,k+3/2}^{n+1/2}+{B_{z}}_{i,j+1,k-1/2}^{n+1/2}-{B_{z}}_{i,j,k-1/2}^{n+1/2}))\nonumber \\
 & - & \frac{c\tau}{\Delta z}(b_{y}({B_{y}}_{i,j+1/2,k+1}^{n+1/2}-{B_{y}}_{i,j+1/2,k}^{n+1/2})\nonumber \\
 & + & a_{y}({B_{y}}_{i,j+3/2,k+1}^{n+1/2}-{B_{y}}_{i,j+3/2,k}^{n+1/2}+{B_{y}}_{i,j-1/2,k+1}^{n+1/2}-{B_{y}}_{i,j-1/2,k}^{n+1/2}))\nonumber \\
 & - & 4\pi\tau{j_{x}}_{i,j+1/2,k+1/2}^{n+1/2},\label{Ex3}\\
\nonumber \\
{E_{y}}_{i+1/2,j,k+1/2}^{n+1} & - & {E_{y}}_{i+1/2,j,k+1/2}^{n}=-\frac{c\tau}{\Delta x}(b_{z}({B_{z}}_{i+1,j,k+1/2}^{n+1/2}-{B_{z}}_{i,j,k+1/2}^{n+1/2})\nonumber \\
 & + & a_{z}({B_{z}}_{i+1,j,k+3/2}^{n+1/2}-{B_{z}}_{i,j,k+3/2}^{n+1/2}+{B_{z}}_{i+1,j,k-1/2}^{n+1/2}-{B_{z}}_{i,j,k-1/2}^{n+1/2}))\nonumber \\
 & + & \frac{c\tau}{\Delta z}(b_{x}({B_{x}}_{i+1/2,j,k+1}^{n+1/2}-{B_{x}}_{i+1/2,j,k}^{n+1/2})\nonumber \\
 & + & a_{x}({B_{x}}_{i+3/2,j,k+1}^{n+1/2}-{B_{x}}_{i+3/2,j,k}^{n+1/2}+{B_{x}}_{i-1/2,j,k+1}^{n+1/2}-{B_{x}}_{i-1/2,j,k}^{n+1/2}))\nonumber \\
 & - & 4\pi\tau{j_{y}}_{i+1/2,j,k+1/2}^{n+1/2},\label{Ey3}\\
\nonumber \\
{E_{z}}_{i+1/2,j+1/2,k}^{n+1} & - & {E_{z}}_{i+1/2,j+1/2,k}^{n}=\frac{c\tau}{\Delta x}(b_{y}({B_{y}}_{i+1,j+1/2,k}^{n+1/2}-{B_{y}}_{i,j+1/2,k}^{n+1/2})\nonumber \\
 & + & a_{y}({B_{y}}_{i+1,j+3/2,k}^{n+1/2}-{B_{y}}_{i,j+3/2,k}^{n+1/2}+{B_{y}}_{i+1,j-1/2,k}^{n+1/2}-{B_{y}}_{i,j-1/2,k}^{n+1/2}))\nonumber \\
 & - & \frac{c\tau}{\Delta y}(b_{x}({B_{x}}_{i+1/2,j+1,k}^{n+1/2}-{B_{x}}_{i+1/2,j,k}^{n+1/2})\nonumber \\
 & + & a_{x}({B_{x}}_{i+3/2,j+1,k}^{n+1/2}-{B_{x}}_{i+3/2,j,k}^{n+1/2}+{B_{x}}_{i-1/2,j+1,k}^{n+1/2}-{B_{x}}_{i-1/2,j,k}^{n+1/2}))\nonumber \\
 & - & 4\pi\tau{j_{z}}_{i+1/2,j+1/2,k}^{n+1/2},\label{Ez3}\\
\end{eqnarray}

\noindent where we are using the following expressions for the free
parameters $a_{\alpha},b_{\alpha}$:

\begin{eqnarray}
a_{x} & = & a_{y}+a_{z},\nonumber \\
a_{y} & = & 0.125\frac{\Delta x}{\Delta y},\nonumber \\
a_{z} & = & 0.125\frac{\Delta x}{\Delta z},\nonumber \\
b_{x} & = & 1-2a_{x},\nonumber \\
b_{x} & = & 1-2a_{y},\nonumber \\
b_{x} & = & 1-2a_{z}.\label{coef}
\end{eqnarray}

\noindent Here we have chosen the coefficients (\ref{coef}) in such
a way, that the scheme is stable provided $\Delta x\le\Delta y,\Delta z$,
and the numerical dispersion is removed for waves running along $X-$axis.

\section{Lorentz boost}

\begin{figure}
\centerline{\includegraphics[width=0.9\textwidth]{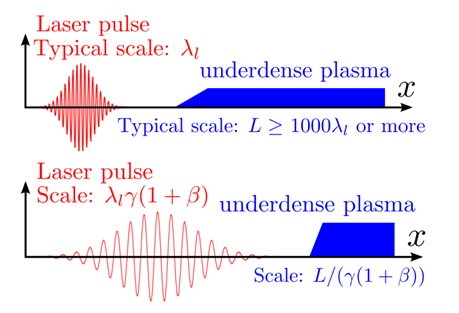}} 
\caption{
Scales discrepancy in plasma-based acceleration. Transformation into
a co-moving frame reduces the scales discrepancy.}

\label{fig:scales-boost} 
\end{figure}

The plasma-based particle acceleration is a multi-scale problem, and
the scales are very disparate, see Fig. \ref{fig:scales-boost}. The
smallest scale is the laser wavelength $\lambda$ in the case of the
laser-drieven acceleration, or the plasma wavelength $\lambda_{p}$
for beam-driven plasma wake fields. The laser wavelength is on the
order of micrometers, while the plasma wavelength can be from tens
of micrometers to millimeters. The medium scale is the driver length.
It can be comparable to the plasma wavelength in the case of the bubble
\cite{Bubble} and blowout \cite{Blowout} regime, or be much longer
when we rely on self-modulation in plasma \cite{Self-modulation1,SM andreev,SM aNTONSEN,SM beam}.
The largest scale is the acceleration length that can range from centimeters
to hundereds of meters or even kilometers \cite{Esarey}. It is this
discrepancy between the driver scale and the acceleration distance
that makes the simulations rather expensive. 

One of the possibilities to bring the scales together is to change
the reference frame form the laboratory one to the co-moving with
the driver. This is called the Lorentz boost technique \cite{Vay Boost}.
Let us assume, we transform from the laboratory frame $L$ into a frame
$R$ moving in the propagation direction of the driver. The relative
velocity of the $R-$frame is $V=\beta c$ and its relativistic factor
is $\gamma=1/\sqrt{1-\beta^{2}}$. Then, the driver is Lorentz-stretched
in the $R-$frame with the factor $\gamma(1+\beta)$ and the propagation
length is compressed with the same factor, see Fig. \ref{fig:scales-boost}.
Thus, potentially, the Lorentz-transformation allows us to increase
the longitudinal grid step and time step - provided it is the grid
step that limits the time step - and we have smaller distance to propagate.
The overall increase in performace could be huge. 

\begin{figure}
\centerline{\includegraphics[width=0.9\textwidth]{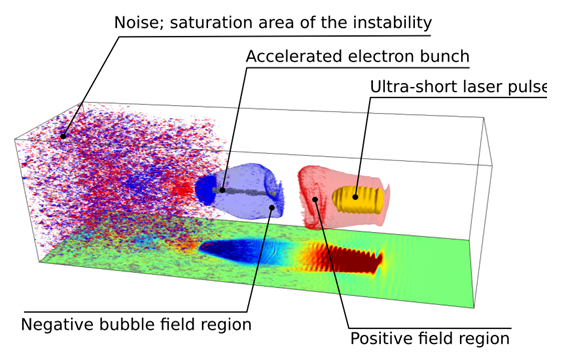}}

\caption{3D view of a numerical instability caused by plasma streaming in the
co-moving frame.}

\label{fig:lorentz-instability-1} 
\end{figure}

Unfortunately, the background plasma becomes streaming in the $R-$frame
with the same transformation velocity $-V$. The plasma density is
higher in the $R-$frame than in the laboratory frame by the factor
$\gamma$. This leads to a source of free energy that can be converted
into numerical plasma heating as the plasma particles interact with
the numerical spatial grid. The associated numerical instability can
have a rather effect on the simulation quality \cite{Vay instab}.
The numerical noise generated by the unstable modes can completely
mask the regular wake structure as shown in the example in Fig. \ref{fig:lorentz-instability-1}. 

\begin{figure}
\centerline{\includegraphics[width=0.5\textwidth]{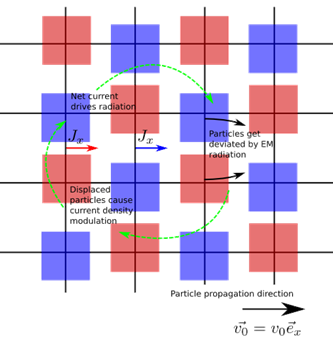}} 

\caption{3D view of a numerical instability caused by plasma streaming in the
co-moving frame.}

\label{fig:lorentz-instability-2} 
\end{figure}

The main reason for the numerical instability is the Cerenkov resonance
between the streaming plasma particles and the numerical electromagnetic
modes on the grid. The numerical electromagnetic modes have subluminal
phase velocities and can be in particle-wave resonance with the macroparticles
of the background plasma that stream through the grid with the relativistic
factor $\gamma$. The mechanism of the numerical instability is illustrated
in Fig. \ref{fig:lorentz-instability-2}a). Plasma fluctuations deviate
particles from the straight line trajectory. This leads to transverse
currents. The transverse currents cause electromagnetic fields. Some
of these electromagnetic modes propagate exactly at the particle velocity
and can resonantly exchange energy with the particles. This is the
Cerenkov mechanism. 

\begin{figure}
\centerline{\includegraphics[width=0.7\textwidth]{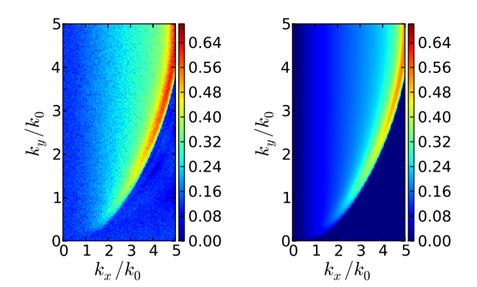}} 

\caption{Left pane: observed growing modes in a PIC simulation, Right pane:
analytically calculated growth rate of Cerenkov-unstable modes.}

\label{fig:lorentz-instability-3} 
\end{figure}

One can solve the wave-particle dispersion relation for the standard
Yee electromagnetic solver and calculate growth rate of the instability
analytically. A comparison of the observed electromagnetic modes in
a PIC simulation and of the analytical prediction is shown in Fig.
\ref{fig:lorentz-instability-3}. Thus, indeed, the reason for the
numerical instability is the Cerenkov resonance between the relativistically
moving plasma particles and numerical electromagnetic modes that have
subluminal phase velocities on the Yee grid.

One might hope to remove the instability by choosing a different Maxwell
solver. One could take for example a dispersion-free solver based
on Fourier transformation. Indeed, a dispersion-free solver reduces
the instability growth rate. Unfortunately, it does not eliminate
it completely. The reason is the spatial and temporal aliasing on
the grid. The relativistic particle starts to interact with the numerical
modes from the other Brillouin zones.

\begin{figure}
\centerline{\includegraphics[width=0.9\textwidth]{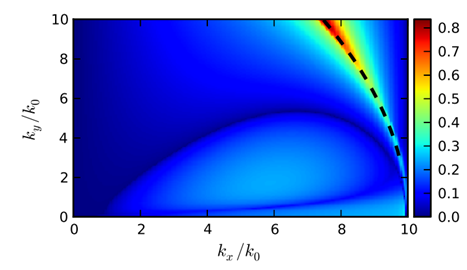}} 

\caption{Observed growing modes in a PIC simulation (the color scale) and the
analytic resonance condition due to the grid aliasing (the broken
line).}

\label{fig:lorentz-instability-NDF} 
\end{figure}

One can write the non-resolved aliased frequencies of the numerical
grid as

\begin{equation}
\omega_{eff}=\pm\left(\sqrt{k_{x}^{2}+k_{y}^{2}}-\frac{1}{\Delta t}\right)\label{eq:aliased freq}
\end{equation}
where $\Delta t$ is the time step. The resonance condition is then
fulfilled for electromagnetic waves with the wave numbers

\begin{equation}
k_{y}\left(k_{x}\right)=\frac{1}{h}\sqrt{h^{2}k_{x}^{2}\left(v_{0}^{2}-1\right)-2hk_{x}v_{0}^{2}+v_{0}^{2}}\label{eq:aliased resonance}
\end{equation}
where $h$ is the grid step. The analytic resonance curve and the
growing modes observed in a numerical experiment are shown in Fig.
\ref{fig:lorentz-instability-NDF}.

The only viable way of taming the numerical instability in Lorentz-boosted
simulations is applying low-pass filters to the deposited currents
before the electromagnetic fields are updated at every time step.
The filtering reduces the instability to acceptable levels. Yet, a
heavy filtering of the currents can in turn influence dispersion of
numerical modes so that one has to be very careful \cite{Vay suppression}.

\section{Quasi-static codes}

Another possibility to bridge the gap between disparate scales in
plasma-based acceleration is using the quasi-static approximation.
It separates explicitly the fast scale of the driver and the slow
scale of acceleration \cite{Wake}. To do so, we introduce the new
variables

\begin{eqnarray}
\tau & = & t\label{eq:tau}\\
\zeta & = & z-ct\label{eq:zeta}
\end{eqnarray}
We assume that the driver changes slowly as it passes its own length.
Thus, as we calculate plasma response to the driver, we neglect all
derivatives over the slow time $\tau$ and we advance from the front
of the driver to the tail to calculate the wake field configuration
at a particular time $\tau$. As we find the fields and the plasma
particles distribution, we can advance the driver with a large time
step in $\tau$. This procedure increases the code performance by
many orders of magnitude. Simulations of large scale plasma-based
acceleration that required huge massively parallel computers with
the explicit PIC, can be done on a desktop workstation in the quasi-static
approximation. Of course, the quasi-static approximation is limited
as it does not describe radiation, but the static electromagnetic
fields only.

The first quasi-static particle-in-cell code WAKE has been written
by Mora and Antonsen \cite{Wake}. It was a 2D code in cylindrical
geometry and use equations written in terms of the wake potential and
the magnetic field. Later, a full 3D code Quick-PIC has been developed
that used the same equations\cite{QuickPIC}. Another 2D code in cylindrical
gemetry LCODE has been developed by Lotov and used equations on fields
directly \cite{LCODE}. Here we show formalism used by the quasi-static
version of the code VLPL. It is a full 3D code in Cartesian geometry.
Like the LCODE, it uses equations on the fields. Below we derive the
quasi-static field equations.

We start again with the Maxwell equations \eqref{Ampere}-\eqref{Bcharge}
and write them in the new variables \eqref{eq:tau}-\eqref{eq:zeta}
neglecting derivatives with respect to the slow time $\tau$:

\begin{eqnarray}
c\nabla\times{\bf B} & = & -c \frac{\partial {\bf E}}{\partial \zeta}+4\pi{\bf j},\label{Ampere-1}\\
c\frac{\partial {\bf B}} {\partial \zeta} & = & c\nabla\times{\bf E},\label{Faraday-1}\\
\nabla\cdot{\bf E} & = & 4\pi\rho,\label{Poisson-1}\\
\nabla\cdot{\bf B} & = & 0.\label{Bcharge-1}
\end{eqnarray}
First, we take curl of the Ampere law \eqref{Ampere-1} and $\zeta$-derivative
of the Faraday law \eqref{Faraday-1}. Combining these two equations,
we arrive at the first quasi-static equation on the magnetic field:

\begin{eqnarray}
\nabla_{\perp}^{2}{\bf B} & = & -\frac{4\pi}{c}\nabla\times{\bf j},\label{eq:QS B}
\end{eqnarray}
where the $\nabla_{\perp}$ operator acts on coordinates transversal
to the propagation direction.

Now, we take gradient of the Poisson law \eqref{Poisson-1}. We use
the well-known identity from vector analysis $\nabla\left(\nabla\cdot{\bf E}\right)=\nabla^{2}{\bf E}+\nabla\times\nabla\times{\bf E}$
and obtain for the transverse components of the electric field the
equation

\begin{eqnarray}
\nabla_{\perp}^{2}{\bf E}_{\perp} & =4\pi\left(\nabla_{\perp}\rho-\frac{1}{c}\partial_{\zeta}{\bf j_{\perp}}\right) & ,\label{eq:QS Eperp}
\end{eqnarray}
For the longitudinal electric field component we obtain

\begin{eqnarray}
\nabla_{\perp}^{2}E_{||} & =4\pi\frac{\partial}{\partial {\zeta}}\left(\rho-\frac{1}{c}j_{||}\right) & =\frac{4\pi}{c}\nabla_{\perp}\cdot{\bf j_{\perp}},\label{eq:QS Epar}
\end{eqnarray}
where also used the continuity equation

\begin{eqnarray}
\frac{\partial}{\partial {\zeta}}\rho & =\frac{\partial}{\partial {\zeta}} j_{||}+\nabla_{\perp}\cdot{\bf j_{\perp}}\label{eq:QS cont}
\end{eqnarray}

The continuity equation \eqref{eq:QS cont} can be used to remove
the charge density $\rho$ from the quasi-static equations and work
with the currents only. This may reduce the noise in particle-in-cell
codes.


\begin{figure}
\centerline{\label{fig:quasi-staticcycle}\includegraphics[width=0.9\textwidth]{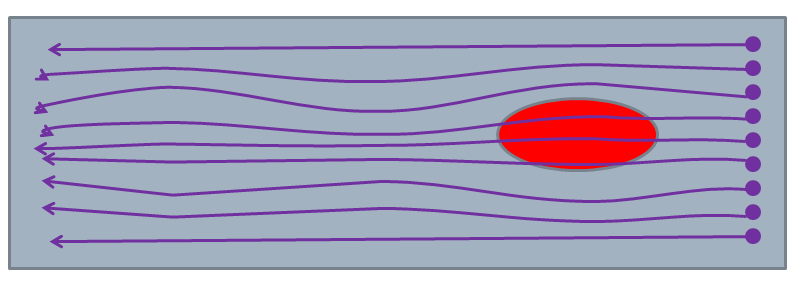}} 

\caption{Left pane: observed growing modes in a PIC simulation, Right pane:
analytically calculated growth rate of Cerenkov-unstable modes.}

\label{fig:quasi-static particles} 
\end{figure}


A typical quasi-static PIC code works a s following. First, the charge
density and currents generated by the driver on the numerical grid
are gathered. These are the sources that contribute to the basic equations
\eqref{eq:QS B}-\eqref{eq:QS Epar}. Then, a layer of numerical macroparticles
is seeded at the front boundary (the head of the driver) of the simulation
box. These numerical particles advance in the negative $\zeta-$direction
(towards the tail of the driver) according to the Eqs. \eqref{eq:QS B}-\eqref{eq:QS Epar}.
As the plasma particles pass the whole simulation domain, the fields
and density are defined on the grid and can be used to advance in
time $\tau$ the driver.

If the driver is a charged particle beam, then we solve equations
of motion for beam particles in the calculated plasma fields. If the
driver is a laser pulse, one has to solve an envelope equation on
the laser pulse amplitude. This is required, because the quasi-static
equations do not describe radiation. Thus, an independent analytical
model is required for the laser pulse. The laser pulse vector potential
is represented as ${\bf A}({\bf \tau,\zeta,r})=\Re\left[{\bf a}({\bf \tau,\zeta,r})\exp\left({\bf i}k\zeta\right)\right]$.
The envelope equation on the complex amplitude ${\bf a}({\bf \tau,\zeta,r})$
reads as \cite{Wake}:

\begin{equation}
\left[\frac{2}{c}\frac{\partial}{\partial\tau}\left({\bf i}k_{0}+\frac{\partial}{\partial\zeta}\right)+\nabla_\perp^{2}\right]{\bf a}=\chi({\bf \zeta,r})\label{eq:schroedinger}
\end{equation}
where $\chi({\bf \zeta,r})=<4\pi q^{2}n/\gamma mc^{2}>$ is the plasma
refraction averaged over all the particles in the cell with the charges
$q$, masses $m$, and relativistic factors $\gamma$.

The laser pulse acts on the plasma particles via its ponderomotive
force

\begin{equation}
F_{p}=-\frac{q^{2}}{\gamma mc^{2}}\nabla a^{2}\label{eq:ponderomotive}
\end{equation}
The ponderomotive force \eqref{eq:ponderomotive} is added to the
standard Lorentz force in the particle pusher.

\begin{figure}
\centerline{\includegraphics[width=0.9\textwidth]{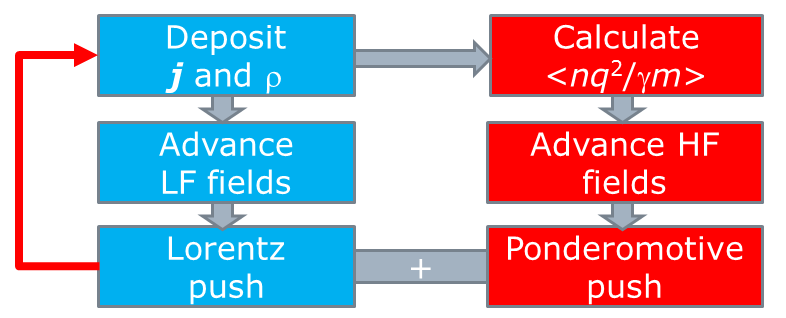}} 

\caption{The typical time step of a quasi-static code. The cycle for low frequency
(LF) plasma fields is followed by the cycle for envelope equation
on the high frequency (HF) laser driver.}

\label{fig:quasi-static principle} 
\end{figure}

The typical time step of a quasi-static code is shown in Fig. \ref{fig:quasi-static principle}.
First, there is a cycle along the fast variable $\zeta$ for the low
frequency (LF) fields. Then, the plasma refraction is calculated and
the envelope equation for the high frequency (HF) fields is updated. 

\begin{figure}
\centerline{\includegraphics[width=0.9\textwidth]{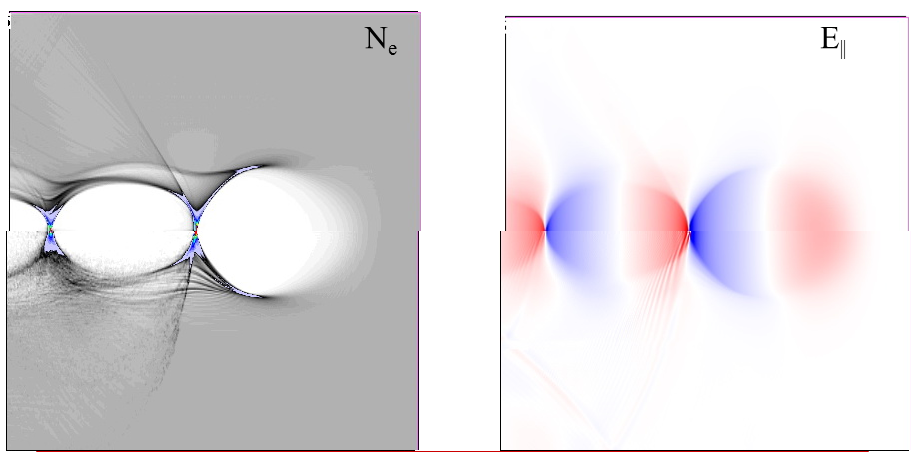}} 

\caption{Blow-out wake field generated by an overdense electron bunch. The
upper half of the frame is taken a quasi-static code simulation. The
lower part is taken from a full PIC code VLPL3D simulation. The left
frame shows the plasma density, the right frame shows the longitudinal
electric field component of the wake field. Cerenkov radiation emitted
by the wave breaking point. is seen in the full 3D PIC simulation.
This radiation is absent in the quasi-static code.}

\label{fig:quasi-static vs PIC 1} 
\end{figure}

Fig. \ref{fig:quasi-static vs PIC 1} shows a comparison between the
quasi-static code (the upper half of the simulation) and the full
PIC code VLPL3D. We have simulated a blow-out generated by an overdense
electron bunch. The electron bunch density had the profile $n_{b}(z,{\bf r})=n_{b0}\exp\left(-z^{2}/2\sigma_{z}^{2}\right)\exp\left(-r^{2}/2\sigma_{r}^{2}\right)$.
The maximum bunch density was two times higher than the background
plasma density: $n_{b0}=2n_{p}$. The bunch was spherical with $k_{p}\sigma_{z}=k_{p}\sigma_{r}=1$.

The both simulations are nearly identical. The only significant difference
is the presence of Cerenkov radiation in the full 3d PIC simulation
from the wave breaking point at the very tail of the first bubble.
In the full simulation, we initialized the electron bunch vacuum in
front of the plasma layer. Thus, the plasma layer had to have a density
ramp. The length of the bubble depends on the plasma density: the
higher the density, the shorter the bubble. Consequently, the wavebreaking
point moved with a superluminal velocity in the density ramp region
and could emit the Cerenkov radiation. In the quasi-static code, the
Cerenkov radiation cannot be simulated. In addition, no density ramp
is needed there: the field distribution is defined by the local plasma
density and by the instantaneous shape of the driver only.

\section{Computational costs of different codes}

It is useful to estimate the number of operations required to simulate
a particular plasma based acceleration problem with different codes.
We call it the computational cost. The most general one, the full
3D PIC code, uses a 3D spatial grid of the size $N_{||}\times N_{\perp}^{2}$
cells and $N_{t}$ time steps. The longitudinal step is limited by
the laser wavelength, $h_{||}\ll\lambda_{0}$ and the corresponding
time step $\tau\ll\lambda_{0}/c$. The transverse grid steps are usually
limited by the plasma wavelength, $h_{\perp}\ll\lambda_{p}$ . The
number of time steps is $N_{t}=L_{acc}/c\tau$, where $L_{acc}$ is
the acceleration distance. Thus, the number of operations required
by the explicit PIC code scales as

\begin{equation}
N_{op}^{PIC}\propto\frac{L_{acc}l_{d}}{\lambda_{0}^{2}}N_{\perp}^{2}\label{eq:Nop PIC}
\end{equation}
Here, we assumed that the ``moving window'' technique is used so
that the longitudinal size of the simulation box $l_{d}$ scales with
the plasma wavelength $\lambda_{p}$.

If one uses the Lorentz boost technique with the transformation relativistic
factor $\gamma_{boost}$, the number of required operations reduces
by the factor $\gamma_{boost}^{2}$ ideally:

\begin{equation}
N_{op}^{PIC\, LB}\propto\gamma_{boost}^{-2}\frac{L_{acc}l_{d}}{\lambda_{0}^{2}}N_{\perp}^{2}=\gamma_{boost}^{-2}N_{op}^{PIC}\label{eq:Nop PIC boost}
\end{equation}

The quasi-static approximation alleviates the time step restriction
by the Courant condition. In addition, the grid step is no more limited
by the laser wavelength $\lambda_{0}$, but rather by the plasma wavelength
$\lambda_{p}$. The time step must resolve the betatron oscillation
of the beam particles, $\tau\omega_{\beta}\ll1$, where the betatron
frequency for a beam particle with the mass $M$ and relativistic
factor $\gamma_{b}$ is $\omega_{\beta}=\omega_{p}\sqrt{m_{e}/2\gamma_{b}M}$.
If the driver is a laser pulse, then the time step must resolve the
diffraction length, $c\tau/Z_{R}\ll1$, where the characteristic diffraction
length of a laser pulse with the focal spot $R$ is defined by the
Rayleigh length $Z_{R}=\pi R^{2}/\lambda_{0}$. The overall number
of operations required by the quasi-static code scales as

\begin{equation}
N_{op}^{QS}\propto\frac{L_{acc}\omega_{\beta}}{c}\frac{l_{d}}{\lambda_{p}}N_{\perp}^{2}=\frac{\lambda_{0}^{2}}{\lambda_{p}\lambda_{\beta}}N_{op}^{PIC}\label{eq:Nop QS}
\end{equation}
where $\lambda_{\beta}=2\pi c/\omega_{\beta}$ is the betatron wavelength.
The performance gain $\lambda_{0}^{2}/\lambda_{p}\lambda_{\beta}$
of the quasi-static codes over the fully explicit PIC can be huge
and easily reach six orders of magnitude.

\section{The future of PIC codes}

Electromagnetic particle-in-cell codes provide a very fundamental
model the dynamics of ideal plasma. Particularly in the relativistic
regime of short pulse laser-plasma interactions, these codes are unique
in the predictive capabilities. In this regime, the binary collisions
of plasma particles are either negligible or can be considered as
a small perturbation and thus the electromagnetic PIC codes are the
most adequate tools.

Yet, the explicit PIC codes have their limits. As soon as one tries
to simulate laser interactions with highly overdense plasmas, these
PIC codes becomes extremely expensive. Indeed, because the scheme
is explicit, the code must resolve the plasma frequency and the skin
depth. Even for uncompressed solid targets of high-Z materials, the
plasma frequency may easily be 30 times higher than the laser frequency.
Correspondingly must be chosen the time and grid steps. For a 3D code,
the simultaneous refinement of the grid and time steps in all dimensions
by a factor $\alpha$ leads to the computational effort increase by
the factor $\alpha^{4}$. For this reason, the simulation of a highly
overdense plasma is still a challenge for the explicit PIC codes.

A way around this difficulty might give implicit PIC codes, like the
code LSP \cite{LSP}, or hybrid, i.e., a mixture of a hydrodynamic
description of the high-density background plasma and PIC module for
the hot electrons and ions \cite{hybrid}. These codes alleviate the
time step limitation, because they suppose the background plasma to
be quasi-neutral and thus eliminate the fastest plasma oscillations
at the Langmuir frequency. Very large plasma regions of high density
can be easily simulated using these codes. Yet, these codes sacrifice
a lot of physics and for each particular problem it must be checked
whether this omitted physics is important or not. One of the possibilities
to do this check is to benchmark the results of implicit codes against
the direct PIC on model problems, which can be simulated by the both
types of the codes.

\section{Aknowledgements}

This work has been supported by EU FP7 project EUCARD-2 and by BMBF, Germany.

\end{document}